\documentclass[aps,prl,twocolumn,groupedaddress,showpacs]{revtex4}

\usepackage{amsmath}
\usepackage{amssymb} 
\usepackage{graphics} 
\usepackage{epsfig}
\usepackage{color}

\allowdisplaybreaks[2]

\newcommand{\beq}{\begin{equation}}
\newcommand{\eeq}{\end{equation}}
\newcommand{\dst}{\displaystyle}
\newcommand{\non}{\nonumber}

\newcommand{\TRUE} {\textrm{\sc true}}
\newcommand{\FALSE}{\textrm{\sc false}}
\newcommand{\Nc}   {{\cal N}}

\newcommand{\muh}  {{\hat{\mu}}}
\newcommand{\vnu}  {{\vec{\nu}}}
\newcommand{\nuh}  {{\hat{\nu}}}
\newcommand{\vlambda}{{\vec{\lambda}}}
\newcommand{\Oc}   {{\cal O}}
\newcommand{\CL}[1][L]{\langle C_{#1}\rangle_N}
\newcommand{\CLG}[1][L]{\langle C_{#1}\rangle^{\trm{G}}_N}
\newcommand{\OL}[1][L]{\langle\Omega_{#1}\rangle_N}
\newcommand{\OLG}[1][L]{\langle\Omega_{#1}\rangle^{\trm{G}}_N}

\newcommand{\OLGinf}[1][L]{\langle\Omega_{#1}\rangle^{\trm{G}}_\infty}
\newcommand{\OLinf}[1][L]{\langle\Omega_{#1}\rangle_\infty}
\newcommand{\rC} {r^\textrm{\scriptsize C}}
\newcommand{\rI} {r^\textrm{\scriptsize I}}

\newcommand{\dr} {\Delta r}
\newcommand{\Nat}{\mathbb{N}}
\newcommand{\real}{\mathbb{R}}

\newcommand{\posi}{\mathbb{Z}_+}
\newcommand{\integers}{\mathbb{Z}}
\newcommand{\trm}{\textrm}
\newcommand{\cNop}{\biggl(1+\frac{\partial_z}N\biggr)^{\!\!N}\bigg|_{z=0}}
\newcommand{\tcNop}{(1+\partial_z/N)^N|_{z=0}}
\newcommand{\cNopdw}{\biggl(1+\frac{\partial_z}N\biggr)^{\!\!N}
                     \partial_w\bigg|_{\substack{z=0\\w=1}}}

\newcommand{\GN}{\langle G\rangle_N}

\newcommand{\gt}{\tilde{g}}
\newcommand{\lambdat}{\tilde{\lambda}}
\newcommand{\avg}[1]{\langle#1\rangle}
\newcommand{\tstir}[2]
           {\bigl[\!\begin{smallmatrix}#1\\#2\end{smallmatrix}\!\bigr]}
\newcommand{\stir}[2]{\begin{bmatrix}#1\\#2\end{bmatrix}}

\newcommand{\hh}{\hat{h}}
\newcommand{\Hh}{\hat{H}}

\begin{document}

\preprint{LU TP 04-43}

\title{Random maps and attractors in random Boolean networks}
%\title{Counting attractors in synchronously updated random Boolean networks}


\author{Bj\"orn Samuelsson}
\email[]{bjorn@thep.lu.se}
%\thanks{}
\author{Carl Troein}
\email[]{carl@thep.lu.se}

\affiliation{
Complex Systems Division, Department of Theoretical Physics\\
Lund University,  S\"olvegatan 14A,  S-223 62 Lund, Sweden}
%\homepage{http://www.thep.lu.se/complex/}


\date{2005-05-07}

\begin{abstract}

Despite their apparent simplicity, random Boolean networks display a
rich variety of dynamical behaviors. Much work has been focused on the
properties and abundance of attractors. The topologies of random
Boolean networks with one input per node can be seen as graphs of
random maps. We introduce an approach to investigating random maps and
finding analytical results for attractors in random Boolean networks with
the corresponding topology. Approximating some other non-chaotic
networks to be of this class, we apply the analytic results to
them. For this approximation, we observe a strikingly good agreement
on the numbers of attractors of various lengths. We also investigate
observables related to the average number of attractors in
relation to the typical number of attractors. Here, we find strong
differences that highlight the difficulties in making direct
comparisons between random Boolean networks and real systems.
Furthermore, we demonstrate the power of our approach by deriving some
results for random maps. These results include the distribution of the
number of components in random maps, along with asymptotic expansions
for cumulants up to the 4th order.

\end{abstract}

%\keywords{}

\maketitle

%\section{}
%\subsection{}
%\subsubsection{}

\section{Introduction}

Random Boolean networks have long enjoyed the attention of
researchers, both in their own right and as simplistic models,
in particular for gene regulatory networks. The
properties of these networks have been studied for a variety
of network architectures, distributions of Boolean rules,
and even for different updating strategies. The simplest and most
commonly used strategy is to synchronously update all nodes.
Networks of this kind have been investigated
extensively, see, e.g., \cite{Kauffman:69, Derrida:86a, Flyvbjerg:88b,
Bastolla:97, Bastolla:98a, Aldana:03, Socolar:03}.

The networks we consider are, generally speaking, such where the
inputs to each node are chosen randomly with equal probability among
all nodes, and where the Boolean rules of the nodes are picked
randomly and independently from some distribution.
In other words, realizing a network of $N$ nodes consists of three steps
to be performed for each node:
(a) choose the number of inputs, called in-degree or connectivity,
and here denoted $K_{\trm{in}}$,
(b) choose a Boolean function of $K_{\trm{in}}$ inputs to be the rule of
the node, and
(c) choose $K_{\trm{in}}$ nodes that will serve as the inputs to the rule.
These steps must be done in the same way for all nodes, and
be independent between nodes. Additionally, though step (c) may be
done with or without replacement, it must give equal
probability to all nodes, implying that the out-degree of each node
is drawn from a Poisson distribution.

The network dynamics under consideration is given by synchronous
updating of the nodes. At any given time step $t$, each node has a
state of $\TRUE$ or $\FALSE$. The state of any node at time
$t+1$ is that which its Boolean rule produces when applied to the
states of the input nodes at time $t$.
Consequently, the entire network state is updated deterministically,
and any trajectory in state space will eventually become periodic.
Thus, the state space consists of attractor basins and
attractors of varying length, and it always has at least one
attractor.

In this work we determine analytically the numbers of attractors of
different lengths in networks with connectivity (in-degree) one. We
compare these results to networks of higher connectivity and find a
remarkable degree of agreement, meaning that networks of single-input nodes
can be employed to approximate more complicated networks, even for
small systems. For large networks, a reasonable level of correspondence
is expected. See \cite{Bastolla:98b} on effective connectivity for
critical networks, and \cite{Kauffman:04} on the limiting numbers of
cycles in subcritical networks.

Random Boolean networks with connectivity one have been investigated
analytically in earlier work \cite{Flyvbjerg:88a, Drossel:05}. In those
papers, a graph-theoretic approach was employed. The approach in
\cite{Flyvbjerg:88a} starts with a derivation that also is directly
applicable to random maps. For a random Boolean
network with connectivity one, a random map can be formed from
the network topology. Every node has a rule that takes its input from
a randomly chosen node. The operation of finding the input node to a
given node forms a map from the set of nodes into itself.  This map
satisfies the properties of a random map.

For highly chaotic networks, with many inputs per node, the state space
can be compared to a random map. Networks where every state is
randomly mapped to a successor state are investigated in
\cite{Derrida:87}.

In \cite{Flyvbjerg:88a}, only attractors with large attractor basins
are considered, and the main results are on the distribution of
attractor basin sizes. We extend these calculations and are able to
also consider attractors with small attractor basins, and include
these in the observables we investigate. \cite{Drossel:05} focuses on
proving superpolynomial scaling, with system size, in the average
number of attractors, as well as in the average attractor length, for
critical networks with in-degree one. Our calculations reveal more
details for cycles of specific lengths.

For long cycles, especially in large networks, there are some
artefacts that make comparisons to real networks difficult. For
example, the integer divisibility of the cycle length is
important, see, e.g., \cite{Flyvbjerg:88a, Bastolla:98b, Samuelsson:03,
Kauffman:04, VKaufman:04}. Also, the total number of attractors grows
superpolynomially with system size in critical networks
\cite{Samuelsson:03, Drossel:05}, and most of the attractors have tiny
attractor basins as compared to the full state space \cite{Bastolla:97,
Bilke:01, Samuelsson:03}. In this work
such artefacts become particularly apparent, and we think that long
cycles are hard to connect to real dynamical systems.

On the other hand, comparisons to real dynamical systems still seem to
be relevant with regard to fixed points and some stability properties
\cite{Kauffman:03, Kauffman:04}. An interesting way to
make more realistic comparisons regarding cycles is to consider those
attractors that are stable with respect to repeated infinitesimal
changes in the timing of updating events \cite{Klemm:04}.

Our approach provides a convenient starting point for
investigations of random maps in general. Random maps have been
the subject of extensive studies, see, e.g., \cite{Kruskal:54, Katz:55,
Harris:60, Ross:81, Jaworski:84, Kupka:90, Donnelly:91, English:93,
Hansen:02}, and also \cite{Bollabas:85} for a book
that includes this subject.
For networks with in-degree one, our approach enables analytical
investigations of far more observables than have been analytically
accessible with previously presented methods. This could provide a
starting point for understanding more complicated networks, and a
tool for seeking observables that may reveal interesting properties in
comparisons to real systems.

Several results on random maps can be
obtained in a straightforward manner from our approach.
One key property of a random map is the number of components
in the functional graph, i.e., the number of separated
islands in the corresponding network. We rederive a relatively simple
expression for the distribution of the number of components, along
with asymptotic expansions for cumulants up to the 4th order. To
a large extent, the asymptotic results are new.

In the results section, we show some numerical comparisons between
random Boolean
networks of multi-input nodes and networks with connectivity one.
The results show similarities that are stronger than we had
expected.
In future research, it is possible that the connection between networks
with single and multiple inputs per node could be better understood by
combining our approach with results and ideas from
\cite{VKaufman:04}. In \cite{VKaufman:04}, the connected Boolean
networks consisting of one two-input node and an arbitrary number of
single-input nodes are investigated.
Although there are difficulties in comparing attractor properties
directly with real dynamical systems, a satisfactory explanation
of the similarities between these networks, with single vs.\ multiple inputs
per node, may provide keys to the understanding of dynamics in
networks in general.

\section{Theory}

In a network with only one input per node, the network topology can be
described as a set of loops with trees of nodes connected to them. To
understand the distribution of attractors of different lengths, it is
sufficient to consider the loops. All nodes outside the loops will
after a short transient time act as slaves to the nodes in the loops.
Also, the nodes in a loop that contains at least one constant rule,
will reach a fixed final state after a short time.

All nodes that are relevant to the attractor structure are contained
in loops with only non-constant (information-conserving) rules.  In
other words, all the relevant elements, as described in
\cite{Bastolla:98a}, are contained in such loops. We let $\mu$ denote
the number of information-conserving loops and let $\muh$ denote the
number of nodes in such loops.

We divide the calculations of the wanted observables into two
steps. First, we present general considerations for loop-dependent
observables. Then, we apply the general results to investigate
observables connected to the attractor structure. Before the second
step, we derive expressions for the distributions of $\mu$ and $\muh$,
together with asymptotic expansions for corresponding means and
variances, to illustrate the meaning and power of the general
expressions.

\subsection{Basic Network Properties}

Throughout this paper, $N$ denotes the number of nodes in the network,
and $L$ the length of an attractor, be it a cycle ($L > 1$) or a fixed
point ($L=1$). For brevity we use the term {\it $L$-cycle}, and
understand this to mean an attractor such that taking $L$ time steps
forward produces the initial state. When $L$ is the smallest positive
integer fulfilling this, we speak of a {\it proper $L$-cycle}.  We
denote the number of proper $L$-cycles, for a given network realization,
by $C_L$. The arithmetic mean over realizations of
networks of a size $N$ is denoted by $\CL$, so the mean number of
network states that are part of a proper $L$-cycle is $L\CL$.

Related to $C_L$ is $\Omega_L$, the number of states that reappear
after $L$ time steps and hence are part of any $L$-cycle, proper or
not. Analogous to $\CL$, we let $\OL$ denote the average of $\Omega_L$
for networks with $N$ nodes. If $\OL$ is known for all $L$, $\CL$ can
be calculated from the set theoretic principle of
inclusion--exclusion. See {\it Supporting Text} to \cite{Kauffman:04}.

For large $N$, the value of $\CL$ is often misleading, in the sense
that some rarely occurring networks with extremely many attractors
dominate the average. To better understand this phenomenon, we
introduce the observables $R_N^L$ and $\OLG$.  $R_N^L$ denotes the
probability that $\Omega_L\ne0$ for a random network of $N$ nodes, and
$\OLG$ is the geometric mean of $\Omega_L$ for $N$-node
networks with $\Omega_L\ne0$.

In the case that every node has one input, the quantities $\OL$,
$R_N^L$ and $\OLG$ can be calculated analytically for any $N$.
In the one-input case, the large-$N$ limit of $\OL$, $\OLinf$, is identical
to the corresponding limit for subcritical networks of multi-input
nodes, as derived in \cite{Kauffman:04}. Furthermore, we discuss in
to what extent critical networks of multi-input nodes
are expected to show similarities to networks of single-input nodes.

For random Boolean networks of one-input nodes, there are only two
relevant control parameters in the model description, apart from the
system size $N$. There are four possible Boolean rules with one
input. These are the constant rules, $\TRUE$ and $\FALSE$, together
with the information-conserving rules that either copy or invert the
input. The distribution of $\TRUE$ vs.\ $\FALSE$ is irrelevant for the
attractor structure of the network. Hence, the relevant control
parameters are the probabilities of selecting inverters and copy
operators when a rule is randomly chosen. We let $\rC$ and $\rI$
denote the selection probabilities associated with copy
operators and inverters, respectively.

In networks with one-input nodes, the total probability of selecting an
information-conserving rule is $r\equiv\rC+\rI$. In analogy with the
definition of $r$, we also define $\dr\equiv\rC-\rI$. In most cases
it is more convenient to work with $r$ and $\dr$ than with $\rC$
and $\rI$. The quantities $r$ and $\dr$ can also be seen as measures
of how a network responds to a small perturbation. From this viewpoint,
$r$ and $\dr$ are average growth factors for a random perturbation
during one time step. For $r$, the size of the perturbation is measured
with the Hamming distance to an unperturbed network. For $\dr$, the
Hamming distance is replaced by the difference in the number of
$\TRUE$ values at the nodes. 

To get suitable perturbation-based definitions of $r$ and $\dr$, we
consider the following procedure:\\
Find the mean field equilibrium fraction of nodes that have the value
$\TRUE$. Pick a random state from this equilibrium as an initial
configuration. Let the system evolve one time step, with and
without first toggling the value of a randomly selected node.
The average fraction of nodes that in both cases copy or invert the
state of the selected node are $\rC$ and $\rI$, respectively.
Finally, let $r=\rC+\rI$ and $\dr=\rC-\rI$. 

It is easy to check that the perturbation-based definitions of $r$
and $\dr$ are consistent with the rule selection probabilities for
networks of single-input nodes.
By using perturbation-based definitions of
$r$ and $\dr$, those quantities are well-defined for networks with
multiple inputs per node \cite{Kauffman:04}, and this allows for direct
comparisons to networks with one input per node.

\subsection{Products of Loop Observables}

In all of our analytical derivations for networks of single-input nodes,
we have a common starting point:\\ We consider observables, on the network,
that can be expressed as a product of observables associated with the
loops in the network.

To make a more precise description, we let $\Nc$ be any
network of single-input nodes, and $\nu$ be the number of loops in
$\Nc$. The dynamical properties of a loop are determined by its
length $\lambda\in\posi$, and a property $s\in\{0,+,-\}$ that we refer to
as the sign of the loop. For a loop that does not conserve
information, i.e., a loop that has at least one constant node,
$s=0$. All other loops have only inverters and copy operators. If the
number of inverters is even then $s=+$, and if it is odd $s=-$.

Let $g_\lambda^s$ denote a quantity that is fully determined by the
length $\lambda$ and the sign $s$ of a loop. We define the
product $G(\Nc)$ of the loop observable $g_\lambda^s$ in $\Nc$
as
\beq
  G(\Nc) \equiv \prod_{i=1}^{\nu}g_{\lambda_i}^{s_i}
\eeq
where $\lambda_1,\ldots,\lambda_\nu$ and $s_1,\ldots,s_\nu$
are the lengths and signs, respectively, of the loops in the network
$\Nc$.

If the network topology is given, but the rules are randomized
independently at each node, the average of $G(\Nc)$ can be calculated
according to
\beq
  \avg{G}_\vlambda = \prod_{i=1}^{\nu}\avg{g}_{\lambda_i}~,
\eeq
where $\vlambda\equiv(\lambda_1,\ldots,\lambda_\nu)$, and
$\avg{g}_\lambda$ is the average of $g_\lambda^s$ under random
choice of rules.

We proceed by also taking the randomization of the network topology
into account. Let $\nu_\lambda$ denote the number of loops of lengths
$\lambda = 1,2,\ldots$, and let $\vnu\equiv(\nu_1,\nu_2,\ldots)$.
Then, the average of $\avg{G}_\vlambda$ over network topologies, in
networks with $N$ nodes, can be written as
\beq
  \avg{G}_N = 
    \sum_{\vnu\in\Nat^\infty}P_N(\vnu)
          \prod_{\lambda=1}^\infty(\avg{g}_\lambda)^{\nu_\lambda}~,
\label{eq: G(P_N)}
\eeq
where $P_N(\vnu)$ is the probability that the distribution of loop
lengths is described by $\vnu$ in a network with $N$ nodes. We use
infinities in the ranges of the sum and the product for formal
convenience. Bear in mind that $P_N(\vnu)$ is nonzero only for
such distributions of loop lengths as are achievable with $N$ nodes.

From \cite{Flyvbjerg:88a}, we know that
\beq
  P_N(\vnu) = \frac\nuh N\frac{N!}{(N-\nuh)!N^\nuh}
    \prod_{\lambda=1}^\infty\frac1{\nu_\lambda!\lambda^{\nu_\lambda}}~,
\label{eq: P_N}
\eeq
where
\beq
  \nuh \equiv \sum_{\lambda=1}^\infty\lambda\nu_\lambda~.
\eeq
Eq.~\eqref{eq: P_N} provides a fundamental starting point for all of
our derivations. In its raw form, however, eq.~\eqref{eq: P_N} is
difficult to work with. In Appendix A we present how to combine
eq.~\eqref{eq: P_N} with eq.~\eqref{eq: G(P_N)}, to obtain
\beq
  \GN = \cNop\exp\sum_{\lambda=1}^\infty
          \frac{\avg{g}_\lambda-1}\lambda z^\lambda~.
\label{eq: GN exp}
\eeq

To continue from eq.~\eqref{eq: GN exp}, we express
$\avg{g}_\lambda$ in terms of more
fundamental quantities. With $\rC$ and $\rI$ as the
probabilities that the rule at any given node is a copy operator or an
inverter, respectively, the probability $p_\lambda^+$ that a loop
of length $\lambda$ has an even number of inverters is given by
\beq
  p_\lambda^+ = \tfrac12[(\rC+\rI)^\lambda+(\rC-\rI)^\lambda]~.
\eeq
Similarly, the probability for an odd number of inverters is given by
\beq
  p_\lambda^- = \tfrac12[(\rC+\rI)^\lambda-(\rC-\rI)^\lambda]~.
\eeq

With $r\equiv\rC+\rI$ and $\dr\equiv\rC-\rI$, we see that
\begin{align}
  p_\lambda^+ &= \tfrac12[r^\lambda+(\dr)^\lambda]~,\\
  p_\lambda^- &= \tfrac12[r^\lambda-(\dr)^\lambda]~,
\intertext{and}
  p_\lambda^0 &= 1-r^\lambda~.
\end{align}

A loop that does not conserve information will always reach a specific
state in a limited number of time steps. Such loops are not relevant
for the attractor properties we are interested in. Thus, $g_\lambda^0$
should not alter the products, and we have $g_\lambda^0=1$. This gives
us
\begin{align}
  \avg{g}_\lambda
    &= p_\lambda^+g_\lambda^+ + p_\lambda^-g_\lambda^-
      +p_\lambda^0g_\lambda^0\\
    &= \gt_\lambda + 1 - r^\lambda~,
\label{eq: g avg r}
\end{align}
where
\beq
  \gt_\lambda = \tfrac12[r^\lambda+(\dr)^\lambda]g_\lambda^+
               +\tfrac12[r^\lambda-(\dr)^\lambda]g_\lambda^-~.
\label{eq: gt}
\eeq

Insertion of eq.~\eqref{eq: g avg r} into eq.~\eqref{eq: GN exp}
and the power series expansion of $\ln(1-x)$ yield
\beq
  \GN = \cNop(1-rz)\exp\sum_{\lambda=1}^\infty
         \frac{\gt_\lambda}\lambda z^\lambda~.
\label{eq: GN exp r}
\eeq
Eq.~\eqref{eq: GN exp r} is the starting point for all network
properties we calculate.

\subsection{Network Topology}

In this section, we investigate the distributions of the number of
information-conserving loops $\mu$ and the number of nodes in those
loops, $\muh$. Both $\mu$ and $\muh$ are independent of whether the
information-conserving loops have positive or negative signs. This
means that $g_\lambda^+=g_\lambda^-$ for all
$\lambda=1,2,\ldots$. Hence, we let $g_\lambda^\pm\equiv
g_\lambda^+=g_\lambda^-$, and get \beq \gt_\lambda = g_\lambda^\pm
r^\lambda~, \eeq which means that eq.~\eqref{eq: GN exp r} turns into
\beq \GN = \cNop(1-rz)\exp\sum_{\lambda=1}^\infty
\frac{g_\lambda^\pm}\lambda (rz)^\lambda~.
\label{eq: GN exp rsym}
\eeq

To investigate the distributions of $\mu$ and $\muh$, we will use
{\it generating functions}. A generating function is a function such
that a desired quantity can be extracted by calculating the
coefficients in a power series expansion.

Let $[w^k]$ denote the operator that extracts the $k$th coefficient in
a power series expansion of a function of $w$. Then, the probabilities
for specific values of $\mu$ and $\muh$, in $N$-node networks, are
given by
\begin{align}
  P_N(\mu=k) &= [w^k]\GN & \trm{if } g_\lambda^\pm&\equiv w
\label{eq: g mu}
\intertext{and}
  P_N(\muh=k) &= [w^k]\GN & \trm{if } g_\lambda^\pm&\equiv w^\lambda~.
\label{eq: g muh}
\end{align}
In eq.~\eqref{eq: g mu}, every loop is counted as one, in powers of $w$,
whereas in eq.~\eqref{eq: g muh}, every node in each loop corresponds
to one factor of $w$.

For probability distributions described by generating functions, there
are convenient ways to extract the statistical moments. Let $m$ denote
$\mu$ or $\muh$. Then, $\avg{m}$ and $\avg{m^2}$ can be calculated
according to
\begin{align}
  \avg m &= \partial_w|_{w=1}\sum_{k=0}^\infty P_N(m=k)w^k\\
         &= \partial_w|_{w=1}\GN
\intertext{and}
  \avg{m^2} &= (1+\partial_w)\partial_w|_{w=1}\sum_{k=0}^\infty P_N(m=k)w^k\\
            &= (1+\partial_w)\partial_w|_{w=1}\GN~.
\label{eq: m2 avg}
\end{align}

Starting from eqs.\ \eqref{eq: g mu}--\eqref{eq: m2 avg}, we derive
some results for $\mu$ and $\muh$. The derivations are presented in
Appendix C. For large $N$, the probability distribution of $\mu$
approaches a Poisson distribution with average $\ln[1/(1-r)]$,
%$\ln\frac1{1-r}$,
whereas the limiting distribution of $\muh$ decays exponentially
as $P(\muh=k)\propto r^k$.

In Appendix C, we also calculate asymptotic expansions for the mean
values and variances of $\mu$ and $\muh$, in the case that $r=1$. The
technique to derive asymptotic expansions for products of loop
observables is presented in Appendix B.

For $r=1$, $\mu$ is equivalent to the number of components in a random
map. Similarly, $\muh$ corresponds to the size of the invariant set in
a random map. The invariant set is the set of all elements that can be
mapped to themselves if the map is iterated a suitable number of
times. Such elements are located on loops in the network graph.

Using the tools in Appendices B and C, one can equally well derive
asymptotic expansions for higher statistical moments as for the mean
and variance. In the results section, we state the leading orders of
the asymptotic expansions for the 3rd and 4th order cumulants to the
distribution of the number of components in a random map.

\subsection{On the Number of States in Attractors}

For a given Boolean network with in-degree one, the number of
states $\Omega_L$ in $L$-cycles can be expressed as a product of loop
observables. If $\Omega_L$ is calculated separately for every loop
in the network, the product of these quantities
gives $\Omega_L$ for the whole network.

Every loop with an even number of inverters and length $\lambda$
can have $2^{\gcd(\lambda,L)}$ states that are repeated
after $L$ timesteps, where $\gcd(a,b)$ denotes the greatest common
divisor of $a$ and $b$.  Hence, such a loop will contribute with the
factor $g_\lambda^+=2^{\gcd(\lambda,L)}$ to the product. Similarly,
for a loop with an odd number of inverters, this factor is
$g_\lambda^-=2^{\gcd(\lambda,L)}$ if $L/\gcd(\lambda,L)$ is even and
$g_\lambda^-=0$ otherwise. The requirement that $L/\gcd(\lambda,L)$ is
even comes from the fact that the state of the loop should be inverted
an even number of times during $L$ timesteps.

The condition that $L/\gcd(\lambda,L)$ is even can be reformulated in
terms of divisibility by powers of 2. Let $\lambdat_L$ denote the the
maximal integer power of 2 such that $\lambdat_L\mid L$, where the
relation $\mid$ means that the number on the left hand side is a
divisor to the number on the right hand side. Then, we get
\begin{align}
  L/\gcd(\lambda,L)\trm{ odd} 
      %&\Leftrightarrow p_2(L)\le p_2(\lambda)\\
      &\Leftrightarrow \lambdat_L\mid\lambda~.
\end{align}

With
\begin{align}
  g_\lambda^+ &= 2^{\gcd(\lambda,L)}
\intertext{and}
  g_\lambda^- &= \left\{\begin{array}{lll}
                2^{\gcd(\lambda,L)} & \trm{if}&\lambdat_L\nmid\lambda\\
                0 & \trm{if}&\lambdat_L\mid\lambda
                  \end{array}\right.
\end{align}
inserted into eq.~\eqref{eq: gt}, we get
\begin{align}
  \gt_\lambda = 2^{\gcd(\lambda,L)}
            \left\{\begin{array}{lll}
                r^\lambda&\trm{if}&\lambdat_L\nmid\lambda\\
                \tfrac12[r^\lambda+(\dr)^\lambda]&
                          \trm{if}&\lambdat_L\mid\lambda
                  \end{array}\right..
\label{eq: gt 2pow}
\end{align}

Now, $\OL$ can be calculated from the insertion of eq.~\eqref{eq: gt 2pow}
into eq.~\eqref{eq: GN exp r}. The arithmetic mean, $\OL$, is, however,
in many cases a bad measure of $\Omega_L$ for a typical network. To
see this, we investigate the geometric mean of $\Omega_L$.

We let $\OLG$ be the geometric mean of nonzero $\Omega_L$, and $R_N^L$ be
the probability that $\Omega_L\ne0$, for networks of size $N$.
The probability distribution of $\log_2\Omega_L$ is generated by
a product of loop observables according to
\begin{align}
  P_N(\log_2\Omega_L=k) &= [w^k]\GN~,
\label{eq: OL distr}
\end{align}
with
\begin{align}
  \gt_\lambda = w^{\gcd(\lambda,L)}
            \left\{\begin{array}{lll}
                r^\lambda&\trm{if}&\lambdat_L\nmid\lambda\\
                \tfrac12[r^\lambda+(\dr)^\lambda]&
                          \trm{if}&\lambdat_L\mid\lambda~.
                  \end{array}\right.
\label{eq: gt wpow}
\end{align}

The probability that $\Omega_L=0$ is not included in eq.~\eqref{eq: OL
distr} for $k\in\Nat$. All other possible values of $\Omega_L$ are
included, and this means that
\begin{align}
  R_N^L &= |_{w=1}\GN~.
%\end{align}
\intertext{
Furthermore, it is clear that
}
%\begin{align}
  R_N^L\log_2\OLG &= R_N^L\avg{\log_2 \Omega_L}_N\\
        &= \partial_w|_{w=1}\GN
%\intertext{and}
%  R_N^L\avg{(\log_2\Omega_L)^2} &= (1+\partial_w)\partial_w|_{w=1}\GN
  ~,
\end{align}
where the average of $\log_2\Omega_L$ is calculated with respect to
networks with $\Omega_L\ne0$.

Insertion of eq.~\eqref{eq: gt wpow} into eq.~\eqref{eq: GN exp r}
yields
\begin{align}
  \GN =&~ \cNop F_L(w,z)~,
\intertext{where}
  F_L(w,z) =&~  (1-rz)
    \exp\sum_{\lambda=1}^\infty
         \frac{w^{\gcd(\lambda,L)}}\lambda r^\lambda z^\lambda
     \non\\&\times\exp\sum_{k=1}^\infty
       \frac{w^{\gcd(k\lambdat_L,L)}}{2k\lambdat_L}
        \bigl[(\dr)^{k\lambdat_L}-r^{k\lambdat_L}\bigr]z^{k\lambdat_L}~,
\label{eq: FL(w,z)}
\end{align}
where $\lambdat_L$ is the largest integer power of 2 that divides $L$.

$F_L$ provides a convenient way to describe our results this far.
We have
\begin{align}
  \OL &= \cNop F_L(2,z)~,
\label{eq: OL FL}\\
  R_N^L &= \cNop F_L(1,z)~,
\label{eq: RL FL}
\intertext{and}
  \OLG &= \exp\Biggl[\frac{\ln2}{R_N^L}\cNopdw F_L(w,z)\Biggr]~.
\label{eq: OLG FL}
\end{align}

Note that $L=0$ can be inserted directly into eq.~\eqref{eq: FL(w,z)}
to investigate the distribution of the total number of states in
attractors. This works because $0$ is divisible by any non-zero
number, and hence $\gcd(\lambda,0)=\lambda$ for all
$\lambda\in\posi$. Insertion of $L=0$ into eq.~\eqref{eq: FL(w,z)},
together with standard power series expansions, yields
\begin{align}
  F_0(w,z) &= \frac{1-rz}{1-rwz}~.
\label{eq: F0}
\end{align}

Eq.~\eqref{eq: F0} gives $F_0(1,z)=1$, which means that $R_N^0=1$. The
result $R_N^0=1$ is easily understood, because every network must have
at least one attractor, and thus a nonzero $\Omega_0$.

The limits $\OLinf$, $R_\infty^L$, and $\OLGinf$ of $\OL$,
$R_N^L$, and $\OLG$ as $N\rightarrow\infty$ are in many cases
easy to extract. For power series of $z$ with convergence radii
larger than $1$, we have the operator relation
\begin{align}
  \lim_{N\rightarrow\infty}\cNop &= \bigg|_{z=1}~, 
\end{align}
which means that the limit can be extracted by letting $z=1$ in the
given function. In the cases that fulfill the convergence criterion
above, we get
\begin{align}
  \OLinf &= F_L(2,1)~,
\label{eq: OLinf FL}\\
  R_\infty^L &= F_L(1,1)~,
\label{eq: RLinf FL}
\intertext{and}
  \OLGinf &= \exp[\ln2\,
                  \partial_w|_{w=1}\!\ln F_L(w,1)]~.
\label{eq: OLGinf FL}
\end{align}
With one exception, all of eqs.\ \eqref{eq: OLinf FL}--\eqref{eq:
OLGinf FL} hold if $r<1$. The exception is that eq.~\eqref{eq: OLinf
FL} does not hold if $L=0$ and $r\ge1/2$.

Using the tools in Appendix B, we find that
\begin{align}
  \OL[0] &\approx \left\{\begin{array}{ll}
             \vphantom{\bigg|}\dst
                  \frac{1-r}{1-2r}&\trm{for }r<\tfrac12\\
             \vphantom{\bigg|}
                 \frac12\sqrt{\frac\pi2N}\dst&\trm{for }r=\tfrac12\\
             \vphantom{\big|}
                 \sqrt{\frac\pi2N}e^{[\ln2r-1+1/(2r)]N}\dst&
                                           \trm{for }r>\tfrac12
             \end{array}\right.
\label{eq: OL0 high N}
\intertext{and}
  \OLG[0] &\approx \left\{\begin{array}{ll}
                  2^{r/(1-r)}&\trm{for }r<1\\
                  2^{\sqrt{\pi N/2}}&\trm{for }r=1~,
             \end{array}\right.
\label{eq: OLG0 high N}
\end{align}
for large $N$.  Note that the the leading term in the asymptote of
$\OL[0]$ for $r>1/2$ comes from the pole in $F_0(2,z)$ at
$z=1/(2r)$. If $r>1/2$, then $z=1/(2r)$ lies inside the contour
$|z-1/3|=2/3$, which is used as integration path in Appendix B. See
Appendices C and D for examples on how to use the technique presented
in Appendix B.

Only if $r<1/2$ do $\OL[0]$ and $\OLG[0]$ have the same qualitative
behavior for large $N$. Otherwise the broad tail in the distribution
of $\muh$ dominates the value of $\OL[0]$. If $1/2<r<1$, $\OLG[0]$
approaches a constant, while $\OL[0]$ grows exponentially with
$N$. For the critical case, $r=1$, the qualitative difference lies in
the power of $N$ in the exponent.

For $L\ne0$, the difference between $\OL$ and $\OLG$ is less
pronounced. Both $\OL$ and $\OLG$ approach constants as
$N\rightarrow\infty$ if $r<1$, and they both grow like powers of $N$
if $r=1$. It is also worth noting that $R_\infty^L\ne0$ for $r<1$,
whereas $R_\infty^L=0$ if $r=1$ but $\dr < 1$. In the latter case,
$R_N^L$ approaches $0$ like $N^{-1/(4\lambdat_L)}$; see Appendix D. If
$r=1$ and $\dr=1$, i.e., the network has only copy operators,
$R_N^L=1$ for all $N\in\posi$.

In Appendix D, we investigate $\OL$ and $\OLG$, for $L>0$, in detail
for the case that $r=1$ and $\dr < 1$, which corresponds to the most
commonly occurring cases of critical networks. For large $N$, we have
the asymptotic relations
\begin{align}
  \OL\propto N^{U_L}
\intertext{for the arithmetic mean of the number of $L$-cycle states,
and}
  \OLG\propto N^{u_L}
\end{align}
for the corresponding geometric mean,
with the exponents $U_L$ and $u_L$
given by eqs.\ \eqref{eq: UL} and \eqref{eq: uL} in Appendix D.
For large $L$, we have
\begin{align}
  U_L &\approx \frac{2^L}{2L}~.
\label{eq: UL approx}
\end{align}

The other exponent, $u_L$, which appears in the scaling of the geometric
mean, is trickier to estimate. However, we derive
an upper bound from $\varphi(\ell)\le\ell$, where $\varphi$ is the
Euler function, as described in Appendix D. From this inequality, combined
with eqs.\ \eqref{eq: uL} and  \eqref{eq: hL}, we find that
\begin{align}
  u_L < \frac{\ln2}2d(L)~,
\end{align}
where $d(L)$ is the number of divisors to $L$. To show that $u_L$ is
not bounded for arbitrary $L$, we let $L=2^m$, where $m\in\Nat$, and
find that $h_L=(m+1)/2$ and
\begin{align}
  u_L = \frac{\ln 2}8(m+1)~.
\label{eq: uL 2-pow}
\end{align}

Although $\OL$ and $\OLG$ share the property that the they grow like
powers of $N$, the values of the powers differ strongly in a
qualitative sense. Yet neither case has an upper limit to
the exponent in the power law. Thus, the observation that the total
number of attractors grows superpolynomially with $N$ is true
not only for the arithmetic mean, but also for the geometric mean.
This is consistent with the derivations in \cite{Drossel:05}, that show
that the typical number of attractors grows faster than polynomially
with $N$.

\section{Results}

Our most important findings are the expression for the expectation
value of products of loop observables on the graph of a random map
[eq.~\eqref{eq: GN exp r}] and the asymptotic expansions for such
quantities. Using these tools, we derive new results on basic
properties of random maps, and on Boolean dynamics on the graph of a
random map. In the latter case, we investigate random Boolean networks
with in-degree one, and compare those to more complicated random Boolean
networks.

\subsection{Random Maps}

For critical random Boolean networks with in-degree one, all loops
conserve information. This is because no constant Boolean rules
are allowed in a critical network. For such a network, the number of
information-conserving loops, $\mu$, is also the number of components
of the network graph. This graph is also the graph of a random map.
Thus, $\mu$ can be seen as the number of components in a random map.
Analogous to the interpretation of $\mu$, the number of nodes
in information-conserving loops, $\muh$, can be seen as the number of
elements in the invariant set of a random map.

We derive the probability distributions of $\mu$ and $\muh$, in a form
that also can be obtained from other approaches \cite{Ross:81,
Kupka:90}. For critical networks, we derive asymptotic expansions for
the means and variances of $\mu$ and $\muh$, and find that
\begin{align}
  \avg\mu =&~ \tfrac12(\ln2N+\gamma)+\tfrac16\sqrt{2\pi}N^{-1/2}+\Oc(N^{-1})~,
\label{eq: avg mu asympt}\\
  \sigma^2(\mu) =&~ \tfrac12(\ln2N+\gamma)-\tfrac18\pi^2+\tfrac16(3-2\ln2)\sqrt{2\pi}N^{-1/2}
                 \non\\&
                   +\Oc(N^{-1})~,
\label{eq: var mu asympt}\\
  \avg\muh =&~ \tfrac12\sqrt{2\pi N}-\tfrac13
                   +\tfrac1{24}\sqrt{2\pi}N^{-1/2}+\Oc(N^{-1})~,
\label{eq: avg muh asympt}
\intertext{and}
  \sigma^2(\muh) =&~ \tfrac12(4-\pi)N-\tfrac16\sqrt{2\pi N}
                   -\tfrac1{36}(3\pi-8)
                 \non\\&
                 +\Oc(N^{-1/2})~,
\label{eq: var muh asympt}
\end{align}
where $N$ is the number of nodes in the network, and $\gamma$ is the
Euler-Mascheroni constant. These expansions converge rapidly to
corresponding exact values, for increasing $N$.

The leading terms $\tfrac12(\ln2N+\gamma)$ of eqs.\ \eqref{eq: avg mu
asympt} and \eqref{eq: var mu asympt} have been derived earlier.
See \cite{Kruskal:54, Stepanov:69, Romero:03} on $\avg\mu$ and
\cite{Stepanov:69, Romero:03} on $\sigma^2(\mu)$. The leading term of
eq.~\eqref{eq: avg muh asympt} is found in \cite{Stepanov:69}.  The
other terms in eqs.\ \eqref{eq: avg mu asympt}--\eqref{eq: var muh
asympt} appear to be new.  Some additional terms are presented in
eqs.\ \eqref{eq: avg mu full}--\eqref{eq: var muh full}.

The technique presented in Appendix B let us also
calculate expansions for cumulants of higher orders.
The leading orders of the 3rd and 4th cumulants for the
distribution of $\mu$ give an interesting hint. Let
$\avg{\mu^3}_{\trm{c}}$ and $\avg{\mu^4}_{\trm{c}}$ denote those
cumulants, respectively. Then, we get
\begin{align}
  \avg{\mu^3}_{\trm{c}} &= \avg\mu+\tfrac74\zeta(3)-\tfrac38\pi^2+\Oc(N^{1/2})
\intertext{and}
  \avg{\mu^4}_{\trm{c}} &= \avg\mu+\tfrac{21}2\zeta(3)
       -\tfrac78\pi^2-\tfrac1{16}\pi^4+\Oc(N^{1/2})~,
\end{align}
where $\zeta(s)$ denotes the Riemann zeta function. All cumulants from
the 1st to the 4th order grow like $\tfrac12\ln N$. One could guess
that all cumulants have this property. If so, the distribution of
$\mu$ is very closely related to a Poisson distribution for large
$N$. (Bear in mind that all cumulants for a Poisson distribution are
equal to the average for the distribution.)

Furthermore, it seems like the process of calculating higher order
cumulants, as well as including more terms in the expansions, can be
fully automated. As far as we know, only a very limited number of
terms, and only for mean values and variances, has been derived in
earlier work.

\subsection{Random Boolean Networks}

\begin{figure}[tbf]
\begin{center}
\epsfig{figure=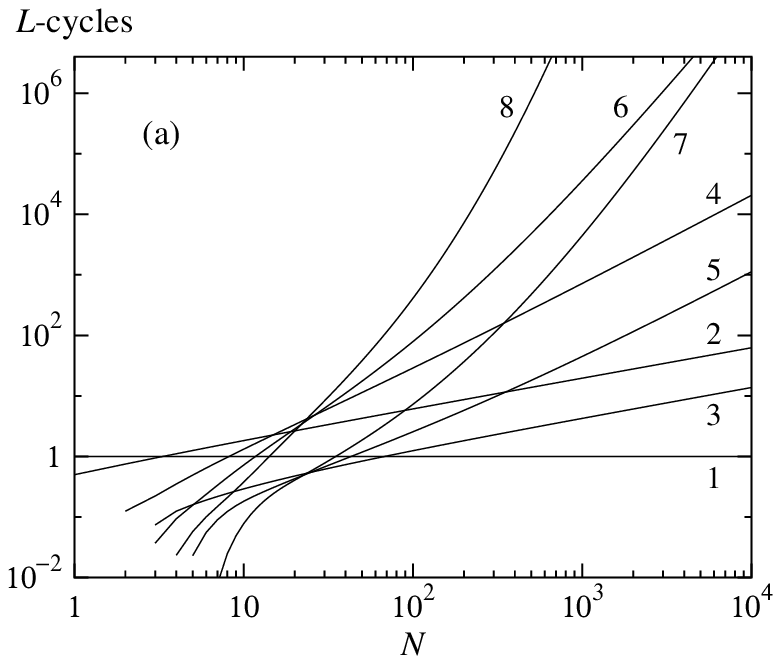}
\epsfig{figure=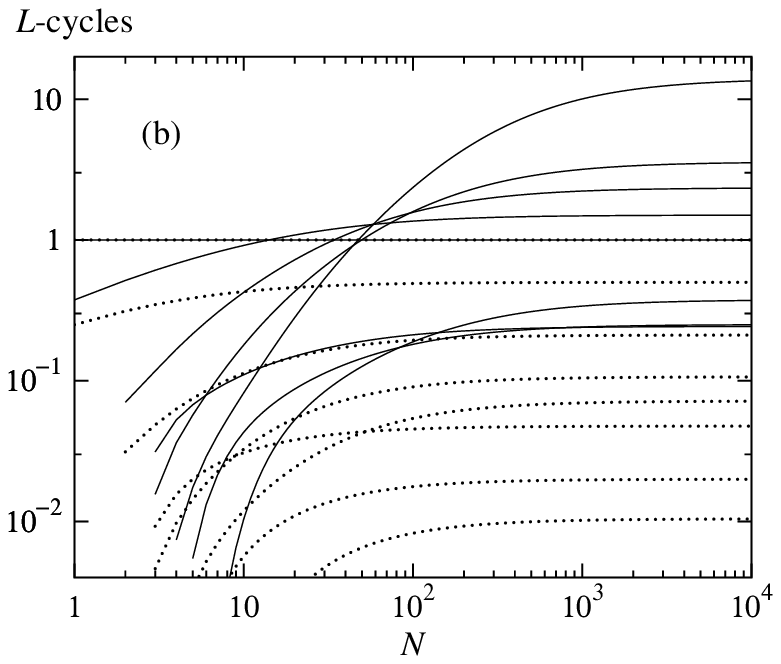}
\end{center}
\caption{The average number of proper $L$-cycles as a function of $N$
for different $L$, for networks with single-input nodes. $r=1$ in (a),
and $r=3/4$ (solid lines) and $r=1/2$ (dotted lines) in (b). In (a),
$L$ is indicated in the plot. In (b), $L$ is 3, 5, 7, 1, 2, 4, 6, and 8
for $r=3/4$ and 7, 5, 3, 8, 6, 4, 2, and 1 for $r=1/2$, in that order,
from bottom to top along the right boundary of the plot area.
In (b), the curves for $L=3$ and $L=5$ for $r=3/4$
essentially coincide at the right side
of the plot, whereas they split up to the left, with $L=3$ as the upper
curve there.}
\label{fig: L cyc sym}
\end{figure}

\begin{figure}[tbf]
\begin{center}
\epsfig{figure=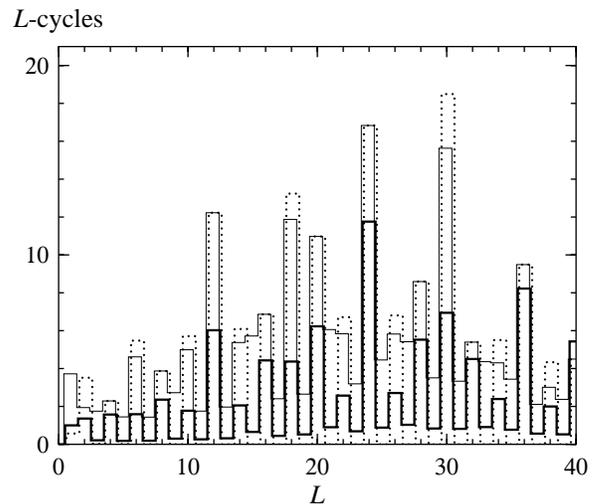}
\end{center}
\caption{The average number of proper $L$-cycles for networks with $N=100$ and
$r=3/4$, as function of $L$. $\dr=3/4$ (thin solid line), $\dr=0$
(thick solid line) and
$\dr=-3/4$ (dotted line). Note the importance of what numbers divide
$L$.}
\label{fig: L cyc asym}
\end{figure}

Our main results from the analytical calculations are the expressions
that yield the arithmetic mean $\OL$, and the geometric mean $\OLG$, of
the number of states in $L$-cycles.
See eqs.\ \eqref{eq: FL(w,z)}--\eqref{eq: OLG FL} on
expressions for general $N$, and eqs.\ \eqref{eq: OLinf
FL}--\eqref{eq: uL 2-pow} on expressions valid for
the high-$N$ limit. In Appendix E, we present derivations that
relate this work to results from
\cite{Kauffman:04}. These derivations yield an expression 
suitable for calculation of exact values of $\OL$ via a power series
expansion of the function $F_L(2,z)$ in eq.~\eqref{eq: OL FL}.

For the arithmetic means, the number of proper $L$-cycles $\CL$ can be
calculated from the number of states $\OL[\ell]$ in all $\ell$-cycles,
provided that $\OL[\ell]$ is known for all $\ell$ that divide
$L$. This is done via the inclusion--exclusion principle as described
in {\it Supporting Text} to \cite{Kauffman:04}. For the corresponding
geometric means we can not use a similar technique, because such
means do not have the needed additive properties.

Our results on random Boolean networks are divided into two parts.
First, we illustrate our quantitative results on networks with
in-degree one. To a large extent, the qualitatively results are expected
from earlier publications.

From \cite{Flyvbjerg:88a}, we know that
in networks with in-degree one, as $N\rightarrow\infty$, the typical number
of relevant variables approaches a constant for subcritical networks, and
scales as $\sqrt{N}$ for critical networks. This
indicates that for subcritical networks, the average number of
$L$-cycles and the average number of states in attractors are
likely to approach constants as $N\rightarrow\infty$.

On the other hand, \cite{Bastolla:97} points out that the probability
distributions of the number of cycles in critical networks have very
broad tails.  Hence, the arithmetic mean can be much larger than the
median of the number of cycles, and this may also be the case for
subcritical networks. In \cite{Kauffman:04}, it is found that this
effect leads to divergence as $N\rightarrow\infty$, in the mean number
of attractors, for networks with the stability parameter $r$ in the range
$r>1/2$. It is also found that the mean number
of cycles of any specific length $L$ converges for large $N$.  For
critical networks, it is clear that both the typical number and the
mean number of attractors grow superpolynomially with $N$, in
networks with in-degree one \cite{VKaufman:04}.

\begin{figure}[tbf]
\begin{center}
\epsfig{figure=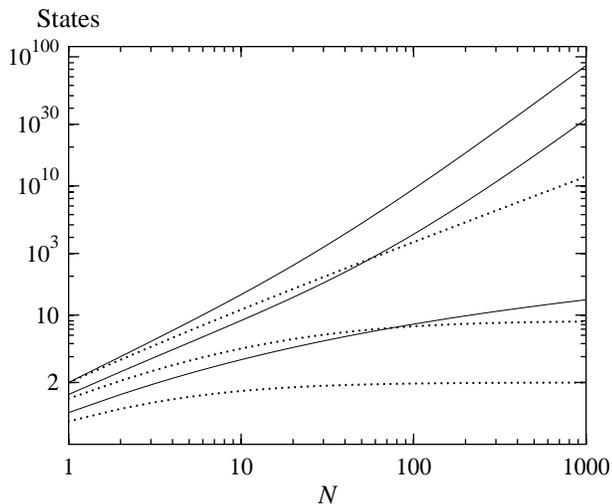}
\end{center}
\caption{Arithmetic and geometric means of the number of states,
$\Omega_0$, in attractors. $\OL[0]$ (solid lines) and $\OLG[0]$
(dotted lines) for $r=1/2$, $r=3/4$ and $r=1$, in that order,
from the bottom to the top of the plot. Note that both $\OL[0]$
and $\OLG[0]$ are independent of $\dr$.}
\label{fig: Omega0 and C}
\end{figure}

Quantitative results that reflect the above properties for networks of
finite sizes are, however, for the most part highly non-trivial to obtain
from earlier work. We let figs.\ \ref{fig: L cyc sym}--\ref{fig: cum
cyc} illustrate our results in this category. Regarding
fig.~\ref{fig: Omega0 and C}, it is important to note that the
geometric mean of the number of states in attractors can be
obtained directly from \cite{Flyvbjerg:88a}.

In the second part of our results on random Boolean networks, we
compare networks with multiple inputs per node to networks with a
single input per node. From a system theoretic viewpoint, this
part is the most interesting, because a general understanding of the
multi-input effects vs.\ single-input effects in dynamical networks
would be very valuable. Although this issue have been addressed before,
in, e.g., \cite{Flyvbjerg:88a,Bastolla:98b}, our results are only partly
explained. These results are illustrated in figs.\ \ref{fig: power vs
1inp}--\ref{fig: crit cut}. 

Fig.~\ref{fig: L cyc sym} shows the numbers of attractors of 
various short lengths as a function of system size, plotted for
different values of the stability parameter $r$. We let $\dr=0$,
corresponding to equal probabilities of inverters and copy operators
in the networks. 
For critical networks, with $r=1$, the asymptotic growth of
the average number of proper $L$-cycles, $\CL$, is a power law,
while $\CL$ approaches a constant for
subcritical networks as $N$ goes to infinity.

For networks with $\dr\ne0$, the prevalences of copy operators and
inverters are not identical. Cycles of even length are in general more
common then cycles of odd length. An overabundance of inverters
strengthens this difference, and conversely a lower fraction of
inverters makes the difference less pronounced. See
fig.~\ref{fig: L cyc asym}, which shows the symmetric case $\dr=0$
and the extreme cases $\dr=\pm r$.

\begin{figure}[tbf]
\begin{center}
\epsfig{figure=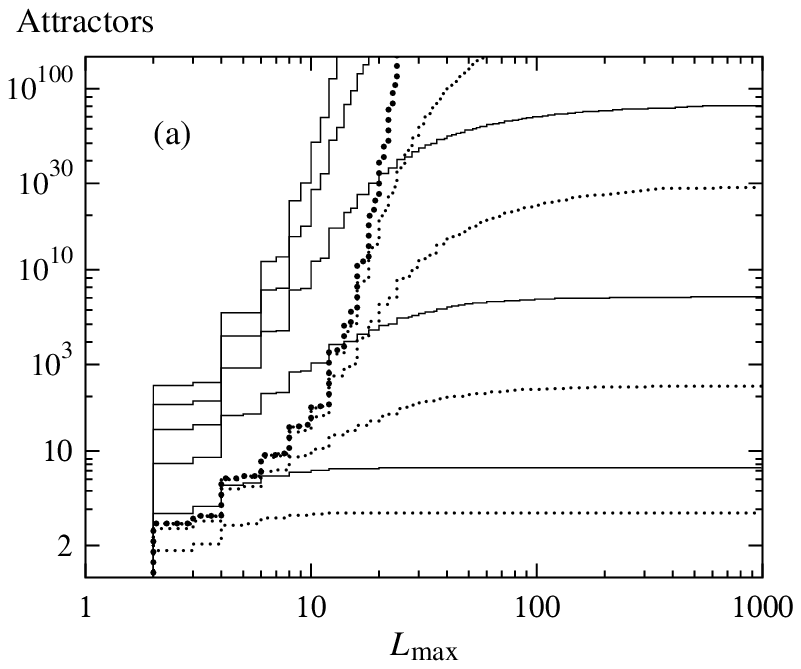}
\epsfig{figure=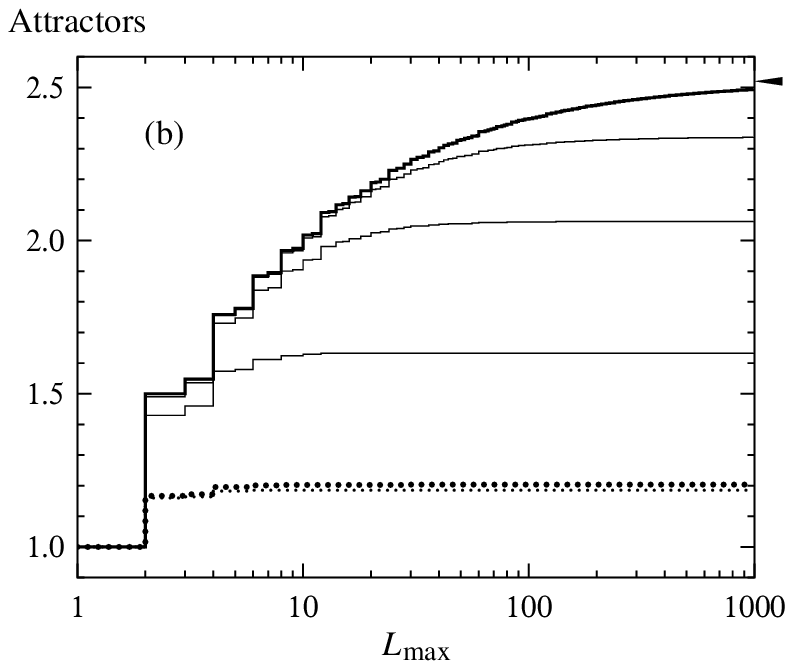}
\end{center}
\caption{The arithmetic mean of the number of attractors with
lengths $L\le L_{\trm{max}}$ in networks with $N$ single-input
nodes, for different values of $N$. In (a) $N=10,10^2,\ldots, 10^5$
for $r=1$ (thin solid lines) and $N=10,\ldots,10^4$ for $r=3/4$ (thin
dotted lines). In (b) $N=10,10^2,10^3$ (thin solid lines) for $r=1/2$
and $N=10$ for $r=1/4$ (thin dotted line). For all cases,
$\dr=0$. The thick lines in (a) and (b) 
show the limiting number of attractors when
$N\rightarrow\infty$. The arrowhead in (b) marks this limit for
$L_{\trm{max}}=10^7$ for $r=1/2$. The small increase in the number of
attractors when $L_{\trm{max}}$ is changed from $10^3$ to $10^7$
indicates that $\CL[]$ converges when $N\rightarrow\infty$. Note the
drastic change in the $y$-scale between the case $r>1/2$ and
$r\le1/2$.}
\label{fig: cum cyc}
\end{figure}

The total number of attractors, $\CL[]$, and the total number of
states in attractors, $\OL[0]$, can diverge for large $N$, even though the
number of attractors of any fixed length converges. This is true for
subcritical networks with $r > 1/2$, and is illustrated in
figs.~\ref{fig: Omega0 and C} and \ref{fig: cum cyc}a.
The growth of $\OL[0]$ is
exponential if $r > 1/2$. Interestingly, there is no qualitative
difference in the growth of $\OL[0]$ when comparing the critical case
of $r = 1$ to the subcritical ones with $1 > r > 1/2$.

For $r < 1/2$, both $\CL[]$ and $\OL[0]$ converge to constants for
large $N$. In the borderline case $r=1/2$, $\OL[0]$ diverges like a
square root of $N$, but $\CL[]$ seems to approach a constant. See
fig.~\ref{fig: cum cyc}b.

The number of states in attractors, $\Omega_0$, of a single-input node
network is directly related to the total number of nodes, $\muh$, that
are part of information-conserving loops. Every state of those nodes
corresponds to a state in an attractor, and vice versa. Thus,
$\Omega_0=2^\muh$, meaning that
\begin{align}
  \OL[0] &= \avg{2^\muh}
\intertext{and}
  \OLG[0] &= 2^{\avg{\muh}}~.
\end{align}

If $1/2<r\leq1$, $\ln\OL[0]$ grows linearly with $N$. This stands in
sharp contrast to $\langle\muh\rangle$, which grows like $\sqrt{N}$
for $r=1$ and approaches a constant for $r<1$ as
$N\rightarrow\infty$. Hence, the distribution of $\muh$ has a broad
tail that dominates $\OL[0]$ if $r>1/2$. This can be understood
from the limit distribution of $\muh$ for large $N$. For this
distribution, we have $P_\infty(\muh=k)\propto r^k$, which means that $r$
must be smaller than $1/2$ for the sum of $2^kP_\infty(\muh=k)$ over $k$
to be convergent. Similar, but less dramatic, effects occur when
forming averages of $\Omega_L$ for $L\ne0$. The arithmetic mean is in
many cases far from the typical value. This is particularly apparent
for long cycles in large networks that are critical or close to
criticality.

In \cite{Drossel:05}, it is shown that the typical number of
attractors grows superpolynomially with $N$ in critical random
Boolean networks with connectivity one. From a different approach, we
find the consistent result that $\CLG[]$ grows superpolynomially,
where $\CLG[]$ is the geometric mean of the number of
attractors. We conclude this from our investigations of the
geometric mean of the number of states in $L$-cycles, $\OLG$.
Here, we use the inequality $\CLG[]\ge\OLG/L$, and the result
that there is no upper bound to $h_L$ in the relation $\OLG\propto
N^{h_L}$, which holds asymptotically for large $N$ (see eq.~\eqref{eq:
uL 2-pow}).

\begin{figure}[tbf]
\begin{center}
\epsfig{figure=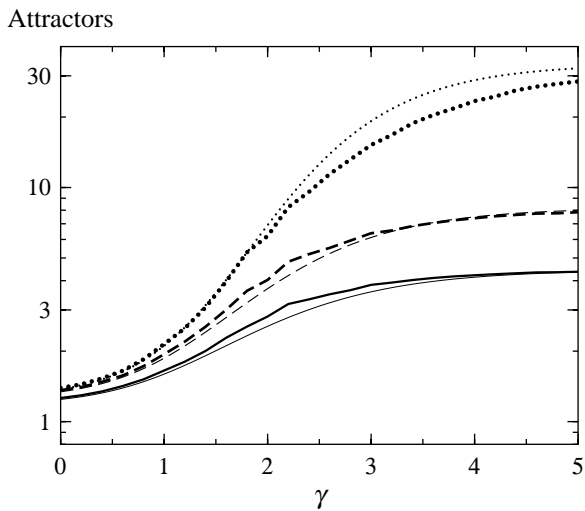}
\end{center}
\caption{Comparison between simulations for power law in-degree
networks of size $N=20$ (bold lines) and the corresponding networks
with single-input nodes (thin lines). The fitted networks have
identical values for $r$, $\dr$, and $N$. The solid lines show the
number of fixed points, whereas the dashed and dotted lines show the
number of 2-cycles plus fixed points and the total number of
attractors, respectively. The probability distribution of in-degrees
satisfies $p_K\propto K^{-\gamma}$, where $K$ is the number of inputs.
The power law networks use the nested
canalyzing rule distribution presented in \cite{Kauffman:04}.}
\label{fig: power vs 1inp}
\end{figure}

\begin{figure}[tbf]
\begin{center}
\epsfig{figure=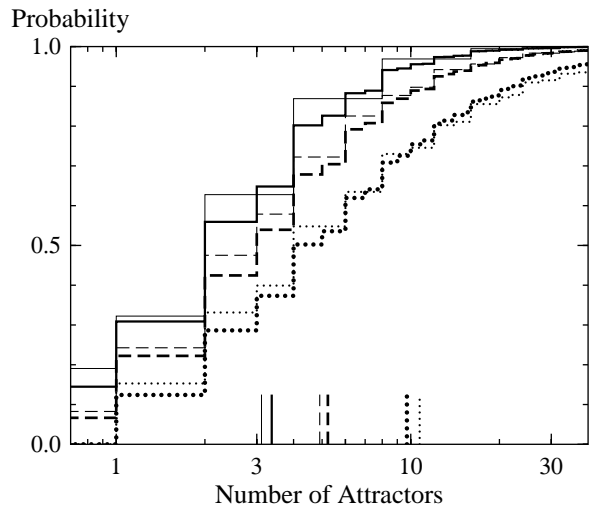}
\end{center}
\caption{A cross-section of fig.~\ref{fig: power vs 1inp} at
$\gamma=2.5$, with simulation results for the power law in-degree
networks (bold lines), and the corresponding single-input networks
(thin lines). The distributions of the number of attractors of
different types are presented with cumulative probabilities, along
with the corresponding means (short vertical lines at the bottom of
the plot). The solid lines show the number of fixed points,
whereas the dashed and dotted lines show the number of 2-cycles plus
fixed points and the total number of attractors, respectively. Note
that the medians are found where the curves for the probability
distributions intersect $1/2$ on the $y$-axis.}
\label{fig: power cut}
\end{figure}

All the properties above are derived and calculated for networks with
one input per node, but they seem to a large extent to be valid for
networks with multi-input nodes.
From \cite{Kauffman:04}, we know that for subcritical networks the limit of
$\CL$ as $N\rightarrow\infty$ is only dependent on $r$ and $\dr$.
Hence, we can expect that $\CL$ for a subcritical network with
multi-input nodes can be approximated with $\CL'$, calculated for a network
with single-input nodes, but with the same $r$ and $\dr$.

For the networks in \cite{Kauffman:04}, with a power law in-degree
distribution, the single-input approximation fits surprisingly
well, which is demonstrated in fig.~\ref{fig: power vs 1inp}. Not only are
the means of the numbers of attractors of different types reproduced by
this approximation, but the distributions of these numbers are also very
similar, as is shown in fig.~\ref{fig: power cut}.

For the critical Kauffman model with in-degree 2, we perform an analogous
comparison. The number of nodes that are non-constant grows
like $N^{2/3}$ for large $N$ \cite{Socolar:03, Samuelsson:03}.
Furthermore, the effective connectivity between the non-constant
nodes approaches 1 for large $N$ \cite{Bastolla:98b}.
Hence, one can expect that this type of $N$-node Kauffman networks can be
mimicked by networks with $N'=N^{2/3}$ one-input nodes. For those
networks, we choose $r=1$ and $\dr=0$, which are the same values as for
the Kauffman networks.

\begin{figure}[tbf]
\begin{center}
\epsfig{figure=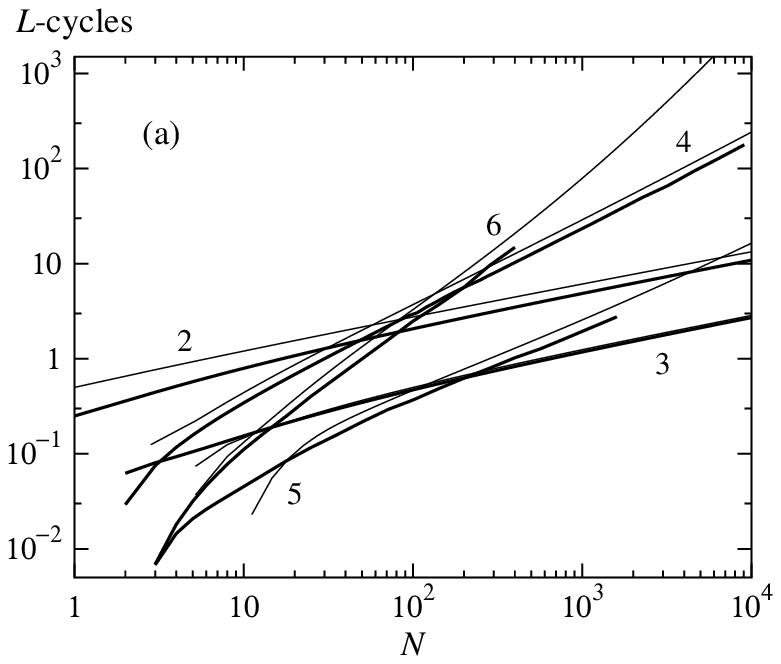}
\epsfig{figure=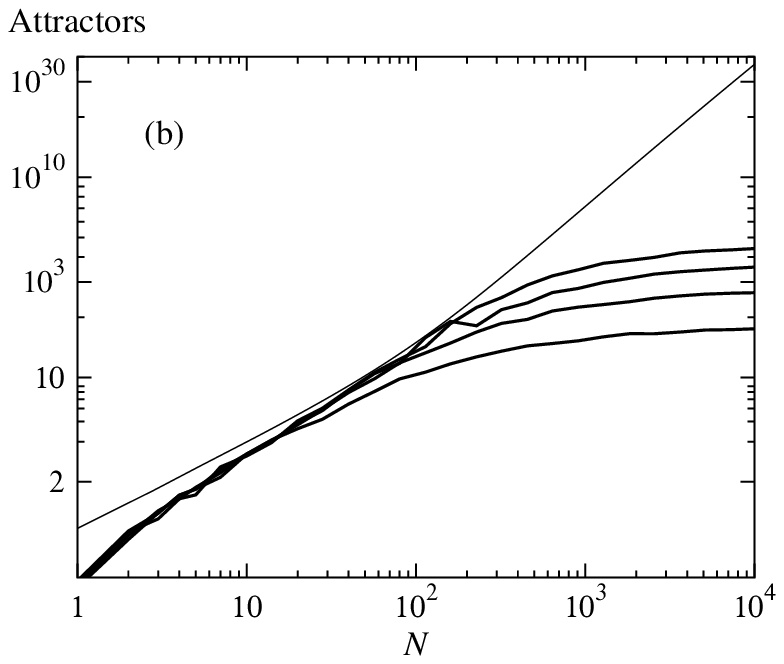}
\end{center}
\caption{Comparison between critical $K=2$ Kauffman networks (thick
lines) and the corresponding networks of single-input nodes (thin
lines). The size of the single-input networks is set to
$N'=N^{2/3}$. $r=1$ and $\dr=0$, consistent with the Kauffman
model. (a) The number of proper $L$-cycles for the $L$ indicated in the plot.
For the Kauffman networks, the
numbers have been calculated from Monte Carlo summation
for those network sizes where could could not be calculated exactly (see
\cite{Samuelsson:03}). The number of fixed points is $1$, independently
of $N$, for both network types. (b) Total
number of attractors. This quantity has been calculated analytically for
the single-input networks, and estimated by simulations
for the Kauffman networks using $10^2$, $10^3$, $10^4$, and $10^5$
random starting configurations per network.}
\label{fig: K2 vs 1inp}
\end{figure}

\begin{figure}[tbf]
\begin{center}
\epsfig{figure=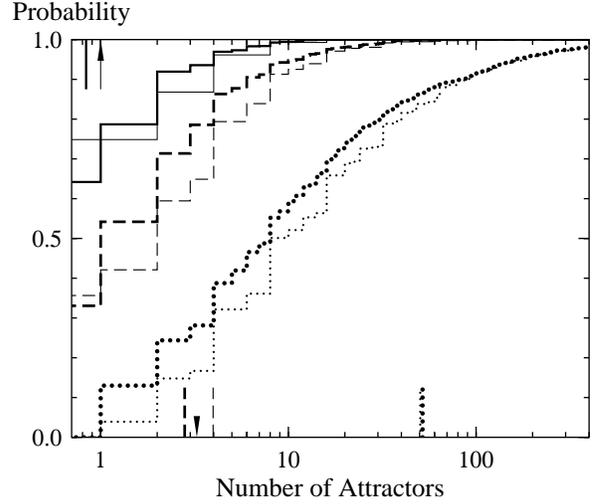}
\end{center}
\caption{A cross-section of fig.~\ref{fig: K2 vs 1inp} at $N=125$,
with simulation results for the Kauffman networks (bold lines), and
the corresponding single-input networks (thin lines). For the Kauffman
networks, we use $10^5$ random starting configurations for $1600$
network realizations. The corresponding single-input node networks have only
$N'=N^{2/3}=25$ nodes, and we perform exhaustive searches through the
state space of the relevant nodes in $10^6$ such networks.
The distributions of the number
of attractors of different types are presented with cumulative
probabilities, along with the corresponding averages (short vertical
lines at the top and bottom of the plot). Corresponding analytical
averages, for the Kauffman networks, are marked with arrowheads.
The solid lines show 
the number of fixed points, whereas the dashed and dotted lines show the
number of 2-cycles plus fixed points and the total number of
attractors, respectively. Note that the medians are found where the curves
for the probability distributions intersect $1/2$ on the $y$-axis.}
\label{fig: crit cut}
\end{figure}

For large $N$, $\CL$ in the Kauffman networks grows like
$N^{(H_L-1)/3}$, where $H_L$ is the number of invariant sets of $L$-cycle
patterns \cite{Samuelsson:03}.
For the selected networks with one-input nodes, we have $\CL'
\propto N'^{(H_L-1)/2} \propto N^{(H_L-1)/3}$ for large $N'$, see
eq.~\eqref{eq: UL}. This confirms that the choice $N'=N^{2/3}$
is reasonable, but it does not indicate whether the proportionality
factor in $N'\propto N^{2/3}$ is anywhere close to 1. This factor could
also be dependent on $L$, as can be seen from the calculations in
\cite{Samuelsson:03}. However, this initial guess turns out to
be surprisingly good, as is shown in fig.~\ref{fig: K2 vs 1inp}a.

From the good agreement for short cycles, one can expect a similar
agreement on the mean of the total number of attractors. This is
investigated in fig.~\ref{fig: K2 vs 1inp}b.
For networks with up to about 100 nodes, the agreement is
good, and the extremely fast growth of $\CL[]'$ for larger $N$ is
consistent with the slow convergence in the simulations.

As with the power law networks, we also compare the distributions
of the numbers of different types of attractors, and find a very
strong correspondence. See fig.~\ref{fig: crit cut}. 
Furthermore, we see indications of undersampling, in the estimated
numbers of fixed points and 2-cycles, for the Kauffman networks in
fig.~\ref{fig: crit cut}, as the means from the simulations are
smaller than the corresponding analytical values.

\section{Summary and Discussion}

Using analytical tools, we have investigated random Boolean networks
with single-input nodes, along with the corresponding random maps.  For
random Boolean networks, we extract the exact distributions of the
average number of cycles with lengths up to 1000 in networks with up
to $10^5$ nodes. As has been pointed out in earlier work
\cite{Bastolla:97}, we see that a small fraction of the networks have
many more cycles than a typical network. This property becomes more
pronounced as the system size grows, and has drastic effects on the
scaling of the average number of states that belong to cycles.

The graph of a random Boolean network of single input nodes can be
seen as a graph of a random map. Our analytical approach is not only
applicable to Boolean dynamics on such a graph, but also to random
maps in general. Using this approach, we rederive some well-known
results in a systematic way, and derive some asymptotic
expansions with significantly more terms than have been available
from earlier publications.
In future research, it would be interesting to, e.g., see to
what extent the ideas from \cite{Hwang:99} and our paper can be
combined.

Our results on random Boolean networks highlight some previously
observed artefacts. The synchronous updates lead to dynamics that
largely is governed by integer divisibility effects. Furthermore,
when counting attractors in large networks, most of them are found in
highly atypical networks and have attractor basins that are extremely
small compared to the full state space. We quantify the role of the
atypical networks
by comparing arithmetic and geometric means of the number of
states in $L$-cycles. From analytical expressions, we find strong
qualitative differences between those types of averages.

The dynamics in random Boolean networks with multi-input nodes can to
a large extent be understood in terms of the simpler single-input case.
In direct comparisons to critical Kauffman networks of in-degree
two and to subcritical networks with power law in-degree, the agreement
is surprisingly good.

In \cite{Klemm:04}, a new concept of stability in attractors of Boolean
networks is presented. To only consider that type of stable attractors
is one way to make more relevant comparisons to real systems. Another
way is to focus on fixed points and stability properties as in
\cite{Kauffman:03} and \cite{Kauffman:04}.
Furthermore, the limit of large systems may not always make
sense in comparison with real systems. Small Boolean networks may tell
more about these than large networks would.

Although there are problems in making direct comparisons between random
Boolean networks and real systems, we think that insight into the
dynamics of Boolean networks will improve the general understanding of
complex systems.  For example, can real systems have lots of
attractors that are never visited due to small attractor basins, and
what implications could such attractors have on the system?

A better understanding of single-input vs.\ multi input dynamics in
Boolean networks could promote a better understanding of similar effects
in more complicated dynamical systems. For the random Boolean networks,
additional insights are required to properly explain
the strong similarities between the single- and multiple-input cases.
One interesting issue is to what extent a single-input approximation can
be applied to networks with random rules on a fixed network graph.

\subsection*{Acknowledgments}

CT acknowledges the support from the Swedish National Research School in 
Genomics and Bioinformatics.

\section{Appendix A: Fundamental Expressions
for Products of Loop Observables}
\renewcommand{\theequation}{A\arabic{equation}}
\setcounter{equation}{0}

Eq.~\eqref{eq: P_N} inserted into eq.~\eqref{eq: G(P_N)} and a
transformation of the summation yield
\begin{align}
  \GN
     &= \sum_{\vnu\in\Nat^\infty}
          \frac\nuh N\frac{N!}{(N-\nuh)!N^\nuh}
          \prod_{\lambda=1}^\infty\frac1{\nu_\lambda!}
             \left(\frac{\avg{g}_\lambda}\lambda\right)
               ^{\nu_\lambda}
\label{eq: GN0}\\
     &= \sum_{\nu=1}^\infty\sum_{\vlambda\in\posi^\nu}
          \frac\nuh N\frac{N!}{(N-\nuh)!N^\nuh}\frac1{\nu!}
          \prod_{i=1}^\nu
             \frac{\avg{g}_{\lambda_i}}{\lambda_i}~.
\label{eq: GN1}
\end{align}
Note that every term in eq.~\eqref{eq: GN0} is split into
$\nu!/\prod_{\lambda=1}^\infty\nu_\lambda!$ equal terms in
eq.~\eqref{eq: GN1}.

Define $c_N^k$ according to
\beq
  c_N^k \equiv \frac{N!}{(N-k)!N^k}~.
\eeq
Then,
\beq
  \GN = \sum_{\nu=1}^\infty\frac1{\nu!}\sum_{\vlambda\in\posi^\nu}
          (c_N^\nuh-c_{N}^{\nuh+1})  
          \prod_{i=1}^\nu
             \frac{\avg{g}_{\lambda_i}}{\lambda_i}~.
\label{eq: GN2}
\eeq
The coefficients $c_N^k$ can be expressed as
\beq
  c_N^k = \cNop z^k~.
\eeq
This relation, together with $\nuh=\sum_{i=1}^\nu\lambda_i$, inserted
into eq.~\eqref{eq: GN2} gives
\begin{align}
  \GN
     &= \cNop\sum_{\nu=1}^\infty\frac1{\nu!}\sum_{\vlambda\in\posi^\nu}
          (1-z)  
          \prod_{i=1}^\nu
             \frac{\avg{g}_{\lambda_i}}{\lambda_i}z^{\lambda_i}
      \non\\
     &= \cNop(1-z)\sum_{\nu=1}^\infty\frac1{\nu!}
        \biggl(\,\sum_{\lambda=1}^\infty
          \frac{\avg{g}_\lambda}\lambda z^\lambda
            \biggr)^{\!\!\nu}.
\label{eq: GN3}
\end{align}

The outer sum in eq.~\eqref{eq: GN3} can
be modified to start from $\nu=0$ without altering the value of the
expression. This property, together with the power series expansions
\begin{align}
  e^x &= \sum_{k=0}^\infty\frac{x^k}{k!}
\label{eq: exp ser}
\intertext{and}
  \ln(1-x) &= -\sum_{k=1}^\infty\frac{x^k}{k}~,
\label{eq: ln ser}
\end{align}
yields that eq.~\eqref{eq: GN3} can be rewritten into eq.~\eqref{eq: GN exp}.

\section{Appendix B: Asymptotes for Products of Loop Observables}
\renewcommand{\theequation}{B\arabic{equation}}
\setcounter{equation}{0}

To calculate eq.~\eqref{eq: GN exp} for large $N$, we investigate the
operator $\tcNop$. Let $f(z)$ be a function that is analytic for
$z\ne1$, such that $|z-1/3|\le2/3$. Furthermore, we assume that $f(z)$
does not have an essential singularity at $z=1$. Then,
\begin{align}
  \cNop f(z) &= \frac{\partial_z^N}{N^N}\bigg|_{z=0}e^{Nz}f(z)\\
             &= \frac{N!}{2\pi i}\oint_{C(\epsilon)}dz
                      \frac{e^{Nz}}{z^{N+1}}f(z)~,
\label{eq: f(z) oint}
\end{align}
where $\epsilon$ is a small positive number, and $C(\epsilon)$ is the
contour of the region where $z$ satisfies $|z-1/3|\le2/3$ and
$|z-1|\ge\epsilon$.

On the curve $C(\epsilon)$, $|e^{Nz}/z^N|$ is maximal close to $z=1$,
where this expression has a saddle point. Thus, the main contribution
to the integral in eq.~\eqref{eq: f(z) oint}, for large $N$, comes
from the vicinity of $z=1$. Contributions from other parts of
$C(\epsilon)$ are suppressed exponentially with $N$.

To find the asymptotic behavior of eq.~\eqref{eq: f(z) oint}, we
perform an expansion of $f(z)$ around $z=1$ with terms of the form
$c[-\ln(1-z)]^m(1-z)^{-a}$, where $a, c\in\real$, and $m\in\Nat$.
Provided that the expansion has a non-zero convergence radius, the
asymptote of eq.~\eqref{eq: f(z) oint} can be determined to any
polynomial order of $N$.

We start at the special case of $f(z)=(1-z)^{-a}$. For non-integral
$a$, $z=1$ is a branch point of $f(z)$. For such $a$ we let $f(z)$ be
real-valued for real $z<1$ and have a cut line at real $z>1$.  For
$N>\max(0,-a)$, we can change the integration path in eq.~\eqref{eq:
f(z) oint}. Let $C'(\epsilon)$ follow the line $\Re(z)=1$ but make a
turn about $z=1$ in the same way as $C(\epsilon)$. Then,
\begin{align}
  \oint_{C(\epsilon)}dz\frac{e^{N(z-1)}}{iz^{N+1}}f(z)
    &=\int_{C'(\epsilon)}dz\frac{e^{N(z-1)}}{iz^{N+1}}f(z)~.
\label{eq: C to C'}
\end{align}
From Stirling's formula \cite{Feller:68},
\begin{align}
  N!=\frac{n^N}{e^N}\sqrt{2\pi N}\exp\biggl[\frac1{12N}+\Oc(N^{-2})\biggr]~,
\end{align}
and eq.~\eqref{eq: C to C'}, we get
\begin{align}
  \cNop\hspace{-24pt}\hspace{24pt}f(z)
      =&~\sqrt{\frac{N}{2\pi}}
               \biggl[1+\frac1{12N}+\Oc(N^{-2})\biggr]\non\\
        &\times\int_{C'(\epsilon)}dz\frac{e^{N(z-1)}}{iz^{N+1}}f(z)~.
\label{eq: cNop f asympt 0}
\end{align}

Around $z=1$, we have $e^{N(z-1)}/z^N\approx\exp[\frac12N(z-1)^2]$.
This approximation can be used as a starting point for a suitable
expansion. To proceed, we note that we can write
\begin{align}
  \sqrt{\frac{N}{2\pi}}
     \int_{C'(\epsilon)}\frac{dz}i\exp\bigl[\tfrac12N(z-1)^2\bigr](1-z)^{-a}
         &=Z(a)N^{a/2}~,
\end{align}
where
\begin{align}
  Z(a) &\equiv \frac{-i}{\sqrt{2\pi}}
       \int_{C'(\epsilon)}dz\exp\bigl[\tfrac12(z-1)^2\bigr](1-z)^{-a}~.
\label{eq: Z def}
\end{align}
From the fast convergence of $\exp[\tfrac12(z-1)^2]$ along $\Re(z)=1$
for large $|z|$, it is clear that $a\mapsto Z(a)$ is well defined and
continuous for all $a$.

With $y=1-z$, we get
\begin{align}
  \hspace{12pt}&\hspace{-12pt}
  \frac{e^{N(z-1)}}{z^{N+1}}(1-z)^{-a} = \non\\
      =&~ \exp\{N[-y-\ln(1-y)]\}\frac{y^{-a}}{1-y}\\
      =&~ \exp\bigl(\tfrac12Ny^2\bigr)y^{-a}
          \bigl[1+y+\tfrac13Ny^3+y^2+\tfrac7{12}Ny^4\non\\
         &+\tfrac1{18}N^2y^6
            +y^3+\tfrac{47}{60}Ny^5+\tfrac{5}{36}N^2y^7+\tfrac1{162}N^3y^9\non\\
         &+\Oc(y^4)+N\Oc(y^6)+N^2\Oc(y^8)+N^3\Oc(y^{10})+\cdots\bigr]~.
\label{eq: saddle y ser}
\end{align}
We insert this result into eq.~\eqref{eq: cNop f asympt 0}, and get
\begin{align}
  &\cNop (1-z)^{-a} = \non\\
  &~~~=  N^{a/2}\Biggl[\sum_{k=0}^3Z_k(a)N^{-k/2}+\Oc(N^{-2})\Biggr]~,
\label{eq: cNop pow asympt}
\end{align}
where
\begin{align}
  Z_0(a) =&~ Z(a)~,
\label{eq: Z0 start}\\
  Z_1(a) =&~ Z(a-1)+\tfrac13Z(a-3)~,\\
  Z_2(a) =&~ \tfrac1{12}Z(a)+Z(a-2)+\tfrac7{12}Z(a-4)+\tfrac1{18}Z(a-6)~,
\intertext{and}
  Z_3(a) =&~ \tfrac1{12}Z(a-1)+\tfrac{37}{36}Z(a-3)
          +\tfrac{47}{60}Z(a-5)\non\\
         &+\tfrac5{36}Z(a-7)+\tfrac1{162}Z(a-9)~.
\label{eq: Z3 start}
\end{align}

Iterated differentiation of eq.~\eqref{eq: cNop pow asympt} with
respect to $a$ gives
\begin{align}
  &\cNop[-\ln(1-z)]^m(1-z)^{-a} = \non\\
    &= N^{a/2}\bigl(\tfrac12\ln N+\partial_a\bigr)^m\Biggl
         [\sum_{k=0}^3Z_k(a)N^{-k/2}+\Oc(N^{-2})\Biggr]~.
\label{eq: cNop pow ln asympt}
\end{align}

It remains for us to calculate $Z(a)$. For $a<1$, eq.~\eqref{eq: Z def} can
be rewritten as
\begin{align}
  Z(a) &= \frac{1}{\sqrt{2\pi}}
       \int_{-\infty}^\infty dx\exp\bigl(-\tfrac12x^2\bigr)(-ix)^{-a}~,
\end{align}
which means that
\begin{align}
  Z(a) &= 2^{-a/2}\pi^{-1/2}\cos\bigl(\tfrac12\pi a\bigr)
                 \Gamma\bigl(\tfrac12-\tfrac12a\bigr)
\label{eq: Z expr cos}
\end{align}
for $a<1$. From eq.~\eqref{eq: Z def} and partial integration, we find
that 
\begin{align}
  Z(a-2)&=(a-1)Z(a)~,
\label{eq: Z recurr}
\end{align}
which is consistent with eq.~\eqref{eq: Z
expr cos}. Hence, eq.~\eqref{eq: Z expr cos} is valid for all $a$,
provided that the right hand side is replaced with an appropriate
limit in case that $a$ is an odd positive integer. The use of the
limit is motivated by the continuity of $Z$.

The recurrence relation in eq.~\eqref{eq: Z recurr} is useful for
expressing $Z_1$, $Z_2$, and $Z_3$ in more convenient forms. Insertion
into eqs.\ \eqref{eq: Z0 start}--\eqref{eq: Z3 start} and
factorization of the obtained polynomials gives
\begin{align}
  Z_0(a) =&~ Z(a)~,\\
  Z_1(a) =&~ \tfrac13(a+1)Z(a-1)~,\\
  Z_2(a) =&~ \tfrac1{36}a(a+2)(2a-1)Z(a)~,
\intertext{and}
  Z_3(a) =&~ \tfrac1{1620}(a+1)(a+3)(10a^2 - 15a - 1)Z(a-1)~.
\end{align}

To express $Z(a)$ in a more convenient form than eq.~\eqref{eq: Z
expr cos}, we use the relations
\begin{align}
  \cos x &= \prod_{k=0}^\infty\Biggl(1-\frac{x^2}
                {\bigl(k+\tfrac12\bigr)^2\pi^2}\Biggr)
\label{eq: prod cos}
\intertext{and}
  \Gamma(x) &= \frac{e^{\gamma x}}x\prod_{k=1}^\infty
          \biggl(1+\frac xk\biggr)e^{x/k}~,
\label{eq: prod Gamma}
\end{align}
where $\gamma$ is Euler-Mascheroni constant.
See, e.g., \cite{Arfken:01} on
eqs.\ \eqref{eq: prod cos} and \eqref{eq: prod Gamma}. We now get
\begin{align}
  Z(a) &= 2^{a/2}e^{a\gamma/2}
           \prod_{k=0}^\infty\biggl(1+\frac a{2k+1}\biggr)
            \exp\biggl(-\frac a{2k+1}\biggr)
\intertext{and}
  Z(a) &= 2^{a/2}\frac{\Gamma\bigl(\tfrac12a\bigr)}{2\Gamma(a)}\\
       &= 2^{a/2}\frac{\bigl(\tfrac12a\bigr)!}{a!}~.
\end{align}

The first and second order derivatives of $Z(a)$ can be expressed
according to
\begin{align}
  Z'(a) &= Z\partial_a\ln Z(a)~
\intertext{and}
  Z''(a) &= Z\partial^2_a\ln Z(a)+Z[\partial_a\ln Z(a)]^2~,
\intertext{with}
 \partial_a\ln Z(a) &= \tfrac12(\ln2+\gamma)
          -\sum_{k=0}^\infty\frac a{(2k+1)(2k+1+a)}~
\intertext{and}
 \partial^2_a\ln Z(a) &=-\sum_{k=0}^\infty\frac 1{(2k+1+a)^2}~.
\end{align}

When the values and derivatives of $Z(a)$ are calculated for $a=0$ and
$a=1$, one can use the recurrence relation, eq.~\eqref{eq: Z recurr}, to
calculate the corresponding properties for any $a\in\integers$.
See, e.g., \cite{Gradshteyn:65} on infinite sums that are useful in those
derivations.

\section{Appendix C: Statistics for Information Conserving Loops}
\renewcommand{\theequation}{C\arabic{equation}}
\setcounter{equation}{0}

Insertion of eq.~\eqref{eq: GN exp rsym} into eqs.~\eqref{eq: g mu} and
\eqref{eq: g muh} gives
\begin{align}
  P_N(\mu=k) &= [w^k]\cNop(1-rz)^{1-w}~
\label{eq: PN mu}
\intertext{and}
  P_N(\muh=k) &= [w^k]\cNop\frac{1-rz}{1-rwz}~.
\label{eq: PN muh}
\end{align}

An alternative form of the probability generating function in
eq.~\eqref{eq: PN mu}, for the special case $r=1$, is presented in
\cite{Hwang:99}. However, this alternative expression is complicated
in comparison to eq.~\eqref{eq: PN mu}, and it is much easier to
extract the probability distribution and corresponding cumulants,
along with their asymptotic expansions, from eq.~\eqref{eq: PN mu}.
In \cite{Hwang:99}, general considerations for probability
generating functions are presented, along with several examples of
such functions. 

For a power series of $z$ with convergence radius larger than $1$, we
have the operator relation
\beq
  \lim_{N\rightarrow\infty}\cNop = \bigg|_{z=1}~,
\eeq
which means the limit can be extracted by inserting $z=1$ in the given
function. In eqs.~\eqref{eq: PN mu} and \eqref{eq: PN muh}, $w$ can be
regarded as an arbitrarily small number, which gives arbitrary large
convergence radii in the corresponding power expansions in $z$. Hence,
the limiting probabilities for large $N$ are given by 
\begin{align}
  P_\infty(\mu=k) &= [w^k](1-r)^{1-w}\\
                  &= (1-r)\frac{[-\ln(1-r)]^k}{k!}~
\label{eq: Pinf mu}
\intertext{and}
  P_\infty(\muh=k) &= [w^k]\frac{1-r}{1-rw}\\
                   &= (1-r)r^k~.
\label{eq: Pinf muh}
\end{align}

Both limiting distributions are normalized for $r<1$, but not for
$r=1$. This means that the probability distributions remains localized
for subcritical networks as $N$ goes to infinity. For critical
networks, the probabilities approach zero, which means that the typical
values of $\nu$ and $\nuh$ must diverge with $N$.

Note that eq.~\eqref{eq: Pinf mu} corresponds to a Poisson distribution
with intensity $\ln[1/(1-r)]$, and that the probabilities in
eq.~\eqref{eq: Pinf muh} decay exponentially with rate $r$. For $\mu$,
we get
\begin{align}
  \avg \mu &= \cNop[-\ln(1-rz)]\\
           &= \sum_{k=1}^N\frac{N!\,r^k}{k(N-k)!N^k}
\label{eq: avg mu}
\intertext{and}
  \avg{\mu^2} &= \cNop\ln(1-rz)[\ln(1-rz)-1]\\
              &= \sum_{k=1}^N\frac{N!\,r^k}{k(N-k)!N^k}
                  \biggl(1+2\sum_{j=1}^{k-1}\frac1j~\biggr)~.
\label{eq: avg mu2}
\end{align}
If $r=1$, $\mu$ can be seen as the number of components in a random
map. For random maps, the result in eq.~\eqref{eq: avg mu} is
well-known and has been derived in several different ways
\cite{Kruskal:54, Harris:60, Ross:81, Kupka:90}. Alternative
derivations of eq.~\eqref{eq: avg mu2} are found in \cite{Ross:81,
Kupka:90}.

The distribution of $\mu$ can be calculated from
eq.~\eqref{eq: PN mu}. To this end, we consider the series expansion
\beq
(1-x)^{-w} = \sum_{n=0}^\infty\frac{x^n}{n!}\sum_{k=0}^n\stir nk w^k~,
\eeq
where $\tstir nk$ are the sign-less Stirling numbers (see, e.g.,
\cite{Berge:71}). Insertion into eq.~\eqref{eq: PN mu} yields
\begin{align}
  P_N(\mu=k) &= \sum_{n=k}^N
         \frac{r^n}{N^n}\binom Nn\biggl(\stir{n-1}{k-1} - \stir{n-1}{k}\biggr)\\
    &= \sum_{n=k}^N
           \frac{r^n}{N^n}\binom Nn\biggl(\stir{n}{k} - n\stir{n-1}{k}\biggr)\\
%    &= \sum_{n=k}^N
%           \frac{r^n}{N^n}\biggl[\binom Nn-\frac{(n+1)r}N\binom N{n+1}\biggr]
%                \stir{n}{k}\\
    &= \sum_{n=k}^N
           \frac{r^n}{N^n}\biggl(1-r+\frac{nr}N\biggr)\binom Nn\stir{n}{k}~.
\label{eq: PN mu expl}
\end{align}

For the number of nodes in information-conserving loops, eq.~\eqref{eq:
PN muh} yields
\begin{align}
  P_N(\muh = k) &= \frac{r^k}{N^k}\biggl(1-r+\frac{kr}N\biggr)\binom Nkk!~.
\label{eq: PN muh expl}
\end{align}
For $r=1$, eq.~\eqref{eq: PN muh expl} is consistent with the
corresponding results on the distribution of the number of invariant
elements in random maps \cite{Harris:60}.

Also, eq.~\eqref{eq: PN muh expl} provides a simpler way to derive
eq.~\eqref{eq: PN mu expl}. It is well known that the probability for
a random permutation of $n$ to have $k$ cycles is given by
$\frac1{n!}\tstir nk$ (see, e.g., \cite{Harris:60}). Consider all nodes
in information-conserving loops of a Boolean network with in-degree 1.
We denote the set of such nodes by $S$. If we randomize the network
topology, under the constraint that $S$ is given, the network graph in
$S$ will also be the graph of a random permutation of the elements in
$S$. Then, every cycle in this permutation corresponds to an
information-conserving loop in the network. In \cite{Stepanov:69,
Kupka:90}, the corresponding observation for random maps was made.

When the network topology is randomized to fit with a given $S$, only
the size $\muh$ of S matters. Thus,
\begin{align}
  P_N(\mu=k\mid\muh=n) &= \frac1{n!}\stir nk~.
\label{eq: PN mu given muh}
\end{align}
Summation over all possible values of $\muh$ gives
\begin{align}
  P_N(\mu = k) = \sum_{n=k}^N P_N(\mu=k\mid\muh=n)P_N(\muh = n)~,
\label{eq: PN mu given muh sum}
\end{align}
which together with eqs.\ \eqref{eq: PN muh expl} and \eqref{eq: PN mu
given muh} provides a simpler derivation of eq.~\eqref{eq: PN mu
expl}. An analogous derivation for random maps is presented in
\cite{Kupka:90}.

For the first and second moments of $\muh$, we find that
\begin{align}
  \avg \muh &= \cNop\frac{rz}{1-rz}\\
            &= \sum_{k=1}^N\frac{N!\,r^k}{(N-k)!N^k}
\label{eq: avg muh}
\intertext{and}
  \avg{\muh^2} &= \cNop\frac{rz(1+rz)}{(1-rz)^2}\\
               &= \sum_{k=1}^N\frac{(2k-1)N!\,r^k}{(N-k)!N^k}~.
\label{eq: avg muh2}
\end{align}

To better understand the results on $\avg\mu$, $\avg{\mu^2}$,
$\avg\mu$, and $\avg{\mu^2}$, we let $r=1$ and calculate their
asymptotes for large $N$. For $r=1$, $\mu$ corresponds to the number of
components in a random map, while $\muh$ corresponds to the number of
elements in its invariant set.

From eq.~\eqref{eq: cNop pow ln asympt}, we find the large-$N$
asymptotes of $\tcNop$ operating on $-\ln(1-z)$, $[\ln(1-z)]^2$,
$(1-z)^{-1}$, and $(1-z)^{-2}$. We also note that $\tcNop1=1$ for all
$N$. From these asymptotes, combined with eqs.\ \eqref{eq: avg mu},
\eqref{eq: avg mu2}, \eqref{eq: avg muh}, and \eqref{eq: avg muh2} for
$r=1$, we get
\begin{align}
  \avg\mu =&~ \tfrac12(\ln2N+\gamma)+\tfrac16\sqrt{2\pi}N^{-1/2}
             -\tfrac1{18}N^{-1}
           \non\\&
             -\tfrac1{1080}\sqrt{2\pi}N^{-3/2}
              +\Oc(N^{-2})~,
\label{eq: avg mu full}\\
  \sigma^2(\mu) =&~ \tfrac12(\ln2N+\gamma)-\tfrac18\pi^2
             +\tfrac16(3-2\ln2)\sqrt{2\pi}N^{-1/2}
           \non\\&
             -\tfrac1{18}(\pi-2)N^{-1}
            -\tfrac1{3240}(41-6\ln2)\sqrt{2\pi}N^{-3/2}
           \non\\&
              +\Oc(N^{-2})~,
\label{eq: var mu full}\\
  \avg\muh =&~ \tfrac12\sqrt{2\pi N}-\tfrac13
                   +\tfrac1{24}\sqrt{2\pi}N^{-1/2}-\tfrac4{135}N^{-1}\non\\
            &+\Oc(N^{-3/2})~,
\intertext{and}
  \sigma^2(\muh) =&~ \tfrac12(4-\pi)N-\tfrac16\sqrt{2\pi N}
                   -\tfrac1{36}(3\pi-8)
                 \non\\&
                   +\tfrac{17}{1080}\sqrt{2\pi}N^{-1/2}
                   +\Oc(N^{-1})~.
\label{eq: var muh full}
\end{align}
Note that the potential term of order $N^{-2}\ln N$ in eq.~\eqref{eq:
var mu full} disappears due to cancellation when $\avg\mu^2$ is
subtracted from $\avg{\mu^2}$.

\section{Appendix D: Asymptotes Related to Boolean Dynamics}
\renewcommand{\theequation}{D\arabic{equation}}
\setcounter{equation}{0}

We take a closer look at the case that $r=1$ and $\dr<1$.
%insertion of eq.~\eqref{eq: RL FL} into
Eq.~\eqref{eq: FL(w,z)} yields
\begin{align}
  F_L(1,z) &=
%  R_N^L  &= \cNop
     \biggl[
       \frac{1-z^{\lambdat_L}}{1-(\dr z)^{\lambdat_L}}
      \biggr]^{1/(2\lambdat_L)}~.
\end{align}
To the leading order in $1-z$, we get
\begin{align}
  F_L(1,z) &=
     \biggl[
       \frac{\lambdat_L(1-z)}{1-(\dr z)^{\lambdat_L}}
      \biggr]^{1/(2\lambdat_L)}[1+\Oc(1-z)]~.
\end{align}
Insertion into eqs.\ \eqref{eq: RL FL} and \eqref{eq: cNop pow
asympt} gives
\begin{align}
  R_N^L =&~ Z\Bigl(\tfrac{-1}{2\lambdat_L}\Bigr)\biggl[
       \frac{\lambdat_L}{1-(\dr z)^{\lambdat_L}}
      \biggr]^{1/(2\lambdat_L)}N^{-1/(4\lambdat_L)}\non\\
        &\times[1+\Oc(N^{-1/2})]~.
\label{eq: RL asympt}
\end{align}

To find the asymptote of $\OLG$, we apply eq.~\eqref{eq: FL(w,z)} and
find that
\begin{align}
  \partial_w|_{w=1}&F_L(w,z)=\non\\
     =&~
     \Biggl(\sum_{\lambda=1}^\infty
            \frac{\gcd(\lambda,L)}{\lambda}z^\lambda\non\\
         &\hphantom{\Biggl(}-\sum_{k=1}^\infty
            \frac{\gcd(k\lambdat_L,L)}{2k\lambdat_L}
            \bigl[1-(\dr)^{k\lambdat_L}\bigr]z^{k\lambdat_L}\Biggr)\non\\
	    &\times\biggl[
       \frac{1-z^{\lambdat_L}}{1-(\dr z)^{\lambdat_L}}
      \biggr]^{1/(2\lambdat_L)}~.
\label{eq: dw1 FL(w,z)}
\end{align}

Let $\varphi$ denote the Euler function. The Euler function is defined
for $n\in\posi$ in such a way that $\varphi(n)$ is the number of
values, $k\in\{1,2,\ldots,n\}$, that satisfy $\gcd(k,n)=1$. If $m$
divides $n$, $\varphi(n/m)$ is the number of values,
$k\in\{1,2,\ldots,n\}$, that satisfy $\gcd(k,n)=m$. From summing over
every $m\in\Nat$ that divides $n$, we get
\begin{align}
  \sum_{m\mid n}\varphi(m/n) &= n~,
\intertext{which means that}
  \sum_{k\mid n}\varphi(k) &= n~.
\label{eq: phi sum}
\end{align}

From eq.~\eqref{eq: phi sum}, we see that
\begin{align}
  \sum_{\lambda=1}^\infty\frac{\gcd(\lambda,L)}{\lambda}z^\lambda
     &=
  -\sum_{\ell\mid L}\frac{\varphi(\ell)}\ell\ln(1-z^\ell)~.
\end{align}
Similarly, we rewrite eq.~\eqref{eq: dw1 FL(w,z)} and get
\begin{align}
  \partial_w|_{w=1}F_L(w,z)=&~%\non\\
     %=&~
     \Biggl[\sum_{\ell\mid L}\frac{\varphi(\ell)}\ell
         \ln\frac1{1-z^\ell}\non\\
      &\hphantom{\Biggl[}\,+\!\sum_{\ell\mid L/\lambdat_L}
          \frac{\varphi(\ell)}{2\ell}
              \ln\frac{1-z^{\ell\lambdat_L}}{1-(\dr z)^{\ell\lambdat_L}}
        \Biggr]\non\\
	    &\times\biggl[
       \frac{1-z^{\lambdat_L}}{1-(\dr z)^{\lambdat_L}}
      \biggr]^{1/(2\lambdat_L)}~.
\label{eq: dw1 FL(w,z) phi}
\end{align}

Again, we perform an expansion around $z=1$ and get
\begin{align}
  \partial_w|_{w=1}F_L(w,z) %= \non\\
        =&~ \bigl[-\hh_L + \tfrac12\bigl(\hh_{L/\lambdat_L} 
                + h_{L/\lambdat_L}\ln\lambdat_L\bigr)\non\\
         &\hphantom{\bigl[} + s_L
              -\bigl(h_L-\tfrac12h_{L/\lambdat_L}\bigr)\ln(1-z)\bigr]\non\\
         &\times\biggl[
             \frac{\lambdat_L}{1-(\dr z)^{\lambdat_L}}
             \biggr]^{1/(2\lambdat_L)}\non\\
         &\times[1+\Oc(1-z)]~,
\end{align}
where
\begin{align}
  h_L &\equiv \sum_{\ell\mid L}\frac{\varphi(\ell)}\ell~,
\label{eq: hL}\\
  \hh_L &\equiv \sum_{\ell\mid L}\frac{\varphi(\ell)}\ell\ln\ell~,
\intertext{and}
  s_L &\equiv 
           \sum_{\ell\mid L/\lambdat_L}
                 \frac{\varphi(\ell)}{2\ell}
              \ln\frac1{1-(\dr z)^{\ell\lambdat_L}}~.
\end{align}

For convenience, we define
\begin{align}
  A_L &= -\hh_L + \tfrac12\bigl(\hh_{L/\lambdat_L} 
                + h_{L/\lambdat_L}\ln\lambdat_L\bigr) + s_L
\intertext{and}
  B_L &= h_L-\tfrac12h_{L/\lambdat_L}~.
\end{align}

Insertion of
\begin{align}
   \partial_w|_{w=1}F_L(w,z)
     &= [A_L - B_L\ln(1-z)][1+\Oc(1-z)]
\end{align}
into eq.~\eqref{eq: OLG FL}, combined with eqs.\ \eqref{eq: cNop pow ln
asympt} and \eqref{eq: RL asympt}, gives
\begin{align}
  \OLG =&~ \exp\Biggl[(\ln2)
           \Biggr(A_L+\tfrac12B_L\ln N
          +B_L\frac{Z'\bigl(\tfrac{-1}{2\lambdat_L}\bigr)}
            {Z\bigl(\tfrac{-1}{2\lambdat_L}\bigr)} \Biggr)
          \Biggr]\non\\
        &\times[1+\Oc(N^{-1/2})]~.
\label{eq: OLG crit high N}
\end{align}
This means that $\OLG$ grows like a power law, $N^{u_L}$, where the
exponent is given by
\begin{align}
  u_L &= \frac{\ln2}2\bigl(h_L - \tfrac12h_{L/\lambdat_L}\bigr)~.
\label{eq: uL}
\end{align}

Finally, we derive the asymptote of $\OL$. From eq.~\eqref{eq:
FL(2,z) old} we get
\begin{align}
  F_L(2,z) &= \frac{S_Le^{-\Hh_L}}{(1-z)^{H_L-1}}[1+\Oc(1-z)]
\end{align}
to the leading order in powers of $1-z$, where
\begin{align}
  H_L &\equiv \sum_{\ell\mid L} J_\ell^++\sum_{2\ell\mid L} J_\ell^-~,\\
  \Hh_L &\equiv \sum_{\ell\mid L}J_\ell^+\ln\ell+\sum_{2\ell\mid L}J_\ell^-\ln\ell~,
\intertext{and}
  S_L &\equiv \prod_{\ell\mid L}\left(\frac1{1-(\dr)^\ell}
                 \right)^{J_\ell^+}
         \prod_{2\ell\mid L}\left(\frac1{1+(\dr)^\ell}
                 \right)^{J_\ell^-}~.
\end{align}
The same procedure as for the other asymptotes lets us find the
asymptote of eq.~\eqref{eq: OL FL}. We obtain
\begin{align}
  \OL &= S_Le^{-\Hh_L}N^{(H_L-1)/2}[1+\Oc(N^{-1})]~.
\label{eq: OL crit high N}
\end{align}
Hence, $\OL$ grows like a power law, $N^{U_L}$, where
\begin{align}
  U_L &= \frac{H_L-1}2~.
\label{eq: UL}
\end{align}
Note that $H_L$ is identical to the number of invariant sets of
$L$-cycle patterns, as defined in \cite{Samuelsson:03}.

\section{Appendix E: An Alternative Expression for $F_L(2,z)$}
\renewcommand{\theequation}{E\arabic{equation}}
\setcounter{equation}{0}

In \cite{Samuelsson:03}, we found that
\begin{align}
  \OLinf =&~ (1-r)
         \prod_{\ell\mid L}\left(\frac1{1-r^\ell}\frac1{1-(\dr)^\ell}
                 \right)^{J_\ell^+}
       \non\\&\times
         \prod_{2\ell\mid L}\left(\frac1{1-r^\ell}\frac1{1+(\dr)^\ell}
                 \right)^{J_\ell^-}~,
\label{eq: OLinf old}
\end{align}
where $J_\ell^\pm$ are integers that can be calculated via the
inclusion--exclusion principle. $J_\ell^-$ satisfies the
relation
\begin{align}
  2\ell J_\ell^-
   &= \sum_{~\mathbf{s}\in\lbrace0,1\rbrace^{\eta_\ell}}
           \!\!
         (-1)^s2^{\ell/d_\ell(\mathbf{s})}~,
\label{eq: J neg}
\end{align}
where $s = \sum_{j=1}^{\eta_\ell} s_j$, $d_\ell(\mathbf{s}) =
\prod_{j=1}^{\eta_\ell} (d_{\ell}^j)^{s_j}$, and
$d_\ell^1,\ldots,d_\ell^{\eta_{\ell}}$ are the odd prime divisors to
$\ell$. Furthermore,
\begin{align}
  J_\ell^+
      &= J_\ell^- - J_{\ell/2}^-~,
\end{align}
where we use the convention that $J_{\ell/2}^-=0$ for odd $\ell$.

From eq.~\eqref{eq: OLinf old}, we can expect that
\begin{align}
  F_L(2,z) =&~ (1-rz)
         \prod_{\ell\mid L}\left(\frac1{1-(rz)^\ell}\frac1{1-(\dr z)^\ell}
                 \right)^{J_\ell^+}
       \non\\&\times
         \prod_{2\ell\mid L}\left(\frac1{1-(rz)^\ell}\frac1{1+(\dr z)^\ell}
                 \right)^{J_\ell^-}~.
\label{eq: FL(2,z) old}
\end{align}
This is indeed true, and to see that, we rewrite eq.~\eqref{eq:
FL(2,z) old} via the power series expansion
\begin{align}
  \ln\frac1{1-x} = \sum_{k=1}^\infty\frac1kx^k~,
\end{align}
and get
\begin{align}
  F_L(2,z) =&~ (1-rz)
         \exp\sum_{\ell\mid L}\sum_{k=1}^\infty \frac{\ell J_\ell^+}{k\ell}
                 [r^{k\ell}+(\dr)^{k\ell}]z^{k\ell}
       \non\\&\times
         \exp\sum_{2\ell\mid L}\sum_{k=1}^\infty \frac{\ell J_\ell^-}{k\ell}
                 [r^{k\ell}+(-1)^k(\dr)^{k\ell}]z^{k\ell}~.
\end{align}
A change of the summation order, with $\lambda=k\ell$, yields
\begin{align}
  F_L(2,z) 
    =&~ (1-rz)
         \exp\sum_{\lambda=1}^\infty\sum_{\ell\mid\gcd(\lambda,L)}
          \hspace{-4pt}
           \frac{\ell J_\ell^+}{\lambda}[r^\lambda+(\dr)^\lambda]z^\lambda
       \non\\&\times
         \exp\sum_{\lambda=1}^\infty
           \sum_{\substack{\ell\mid\gcd(\lambda,L)\\2\mid L/\ell}}
          \hspace{-8pt}
           \frac{\ell J_\ell^-}{\lambda}[r^\lambda+
                  (-1)^{\lambda/\ell}(\dr)^\lambda]z^\lambda
        ~.
\label{eq: FL(2,z) lambda J}
\end{align}
Eq.~\eqref{eq: FL(2,z) lambda J} is consistent with eq.~\eqref{eq:
FL(w,z)}, provided that
\begin{align}
  \sum_{\ell\mid\gcd(\lambda,L)} 
     \hspace{-10pt}
       \ell J_\ell^+ + 
     \hspace{-10pt}
     \sum_{\substack{\ell\mid\gcd(\lambda,L)\\2\mid L/\ell}} 
     \hspace{-10pt}
         \ell J_\ell^-
    &= 2^{\gcd(\lambda,L)}\left\{\begin{array}{lll}
                1 & \trm{if}&\lambdat_L\nmid\lambda\\
                \tfrac12 & \trm{if}&\lambdat_L\mid\lambda~,
                  \end{array}\right.\!
\label{eq: J crit 0a}
\end{align}
and
\begin{align}
  \sum_{\ell\mid\gcd(\lambda,L)}
     \hspace{-10pt}
       \ell J_\ell^+ +
     \hspace{-10pt}
     \sum_{\substack{\ell\mid\gcd(\lambda,L)\\2\mid L/\ell}}
     \hspace{-10pt}
         (-1)^{\lambda/\ell}\ell J_\ell^-
    &= %\non\\ = 
      2^{\gcd(\lambda,L)}\left\{\begin{array}{lll}
                0 & \trm{if}&\lambdat_L\nmid\lambda\\
                \tfrac12 & \trm{if}&\lambdat_L\mid\lambda~.
                  \end{array}\right.\!
\label{eq: J crit 0b}
\end{align}
The sum of eqs.~\eqref{eq: J crit 0a} and \eqref{eq: J crit 0b} is
given by
\begin{align}
  \sum_{\ell\mid\gcd(\lambda,L)} 
     \hspace{-8pt}
       2\ell J_\ell^+ +
     \hspace{-8pt}
     \sum_{2\ell\mid\gcd(\lambda,L)}
     \hspace{-8pt}
       2\ell J_\ell^-
    &= 2^{\gcd(\lambda,L)}~,
\label{eq: J crit sum 0}
\end{align}
which is equivalent to
\begin{align}
  \sum_{\ell\mid\gcd(\lambda,L)} 
     \hspace{-8pt}
       (2\ell J_\ell^+ + \ell J_{\ell/2}^-)
    &= 2^{\gcd(\lambda,L)}
\intertext{and}
  \sum_{\ell\mid\gcd(\lambda,L)} 
     \hspace{-8pt}
       (2\ell J_\ell^- - \ell J_{\ell/2}^-)
    &= 2^{\gcd(\lambda,L)}
     ~.
\label{eq: J crit sum end}
\end{align}
Eq.~\eqref{eq: J crit sum end} is true as a consequence of
eq.~\eqref{eq: J neg}, and hence eq.~\eqref{eq: J crit sum 0} is
true.

The difference between eqs.~\eqref{eq: J crit 0a} and \eqref{eq: J
crit 0b} is
\begin{align}
  \sum_{\substack{\ell\mid\gcd(\lambda,L)\\2\mid L/\ell\\2\nmid\lambda/\ell}}
     \hspace{-8pt}
       2\ell J_\ell^-
    &= 2^{\gcd(\lambda,L)}
         \left\{\begin{array}{lll}
                1 & \trm{if}&\lambdat_L\nmid\lambda\\
                0 & \trm{if}&\lambdat_L\mid\lambda
                  \end{array}\right.
       ~.      
\label{eq: J crit diff 0}
\end{align}
If $\lambdat_L\mid\lambda$, the sum in eq.~\eqref{eq: J crit diff 0}
is empty and therefore equal to the right hand side. If
$\lambdat_L\nmid\lambda$, eq.~\eqref{eq: J crit diff 0} is equivalent
to
\begin{align}
  \sum_{\substack{\ell\mid\gcd(\lambda,L)\\2\ell\nmid\gcd(\lambda,L)}}
     \hspace{-8pt}
       2\ell J_\ell^-
    &= 2^{\gcd(\lambda,L)}
       ~,
\label{eq: J crit diff 1}
\end{align}
consistent with eq.~\eqref{eq: J neg}. Hence, eq.~\eqref{eq: J
crit diff 0} holds, and this result concludes the verification of
eqs.~\eqref{eq: J crit 0a} and \eqref{eq: J crit 0b}. Thus, we
conclude that eq.~\eqref{eq: FL(2,z) old} is correct.

\end{document}